\documentclass[12pt]{article}
\pdfoutput=1
\usepackage{amsmath}
\usepackage{graphicx,color}
\usepackage{amssymb}
\usepackage{srcltx}

\newcommand{\be}{\begin{equation}}
\newcommand{\ba}{\begin{eqnarray}}
\newcommand{\ea}{\end{eqnarray}}
\newcommand{\ee}{\end{equation}}

\newcommand{\bea}{\begin{eqnarray}}
\newcommand{\eea}{\end{eqnarray}}
\newcommand{\bes}{\begin{equation*}}
\newcommand{\beas}{\begin{eqnarray*}}
\newcommand{\eeas}{\end{eqnarray*}}
\newcommand{\bas}{\begin{array*}}
\newcommand{\eas}{\end{array*}}
\newcommand{\ees}{\end{equation*}}

\newcommand*{\mathcolor}{}
\def\mathcolor#1#{\mathcoloraux{#1}}
\newcommand*{\mathcoloraux}[3]{%
  \protect\leavevmode
  \begingroup
    \color#1{#2}#3%
  \endgroup
}

\begin{document}

\begin{titlepage}
\thispagestyle{empty}

\begin{center}
\noindent{\textbf{\large {Topological entanglement negativity in Chern-Simons theories
}}}\\
\date{\today}
\vspace{2cm}
Xueda Wen $^{1}$, Po-Yao Chang $^{2}$, and Shinsei Ryu $^{1}$
\vspace{1cm}

{\it
$^{1}$Institute for Condensed Matter Theory and
Department of Physics, University of Illinois
at Urbana-Champaign,
1110 West Green St, Urbana IL 61801, USA\\
}

{\it
$^{2}$Center for Materials Theory, Rutgers University, Piscataway, NJ, 08854, USA\\
}

\vskip 2em
\end{center}

\begin{abstract}
We study the topological entanglement negativity between two spatial regions in
(2+1)-dimensional Chern-Simons gauge theories by using the replica trick and the surgery method.
For a bipartitioned or tripartitioned spatial manifold,
we show how the topological entanglement negativity
depends on the presence of quasiparticles and the choice of ground states.
In particular, for two adjacent non-contractible regions on a tripartitioned torus,
the entanglement negativity
provides a simple way to distinguish Abelian and non-Abelian theories.
Our method applies to a Chern-Simons gauge theory defined on an arbitrary oriented (2+1)-dimensional
spacetime manifold.
Our results agree with the edge theory approach in
a recent work (X. Wen, S. Matsuura and S. Ryu, arXiv:1603.08534).
\end{abstract}

\end{titlepage}

\tableofcontents

\section{Introduction}

Recently, quantum entanglement provides
a powerful tool to study the properties of quantum many-body systems in condensed matter physics \cite{Kitaev,LevinWen,Calabrese2009,Elsert2010}, 
such as characterizing topological ordered phases, 
and detecting the central charge of conformal field theories, etc.
\cite{Kitaev,LevinWen,Calabrese2009,Elsert2010,Dong,Zhang,Calabrese2004}.

To characterize the quantum entanglement, there are various kinds of entanglement measures.
In the case when a system is prepared in a pure state $|\Psi\rangle$ and bipartitioned into two subsystems $A$ and $B$,
two quantum entanglement measures which turn out to be very useful are the so-called
Renyi entropy and von Neumann entropy defined as follows
\begin{equation}\label{vN}
S_A^{(n)}=\frac{1}{1-n}\ln\text{Tr}\rho_A^n, \quad \text{and} \quad
S_A^{\text{vN}}=-\text{Tr}\rho_A\ln\rho_A,
\end{equation}
where $n$ is an integer,
and $\rho_A=\text{Tr}_B\rho$ is the reduced density matrix of subsystem $A$, with $\rho=|\Psi\rangle\langle\Psi|$.
The Renyi entropy and von Neumann entropy are related by $S_A^{\text{vN}}=\lim_{n\to 1}S_A^{(n)}$.
It is noted that when $\rho$ corresponds to a pure state, one has the nice property that
$S_A^{(n)}=S_B^{(n)}$ and $S_A^{\text{vN}}=S_B^{\text{vN}}$.
For a mixed state, it is found that the quantum and classical correlations cannot be explicitly
separated in these entanglement measures. Now we consider two subsystems $A_1$ and $A_2$
which are embedded in a larger system, and therefore $\rho_{A_1\cup A_2}$ may correspond to a mixed state. In this case, a useful quantity to study the correlation between $A_1$ and $A_2$ is the Renyi mutual information
\begin{equation}
I_{A_1A_2}^{(n)}=S_{A_1}^{(n)}+S_{A_2}^{(n)}-S_{A_1\cup A_2}^{(n)},
\end{equation}
which is symmetric in $A_1$ and $A_2$ by definition.
Similar to the von Neumann entropy, by taking the $n\to 1$ limit, one can obtain
the (von Neumann) mutual information
\begin{equation}
I_{A_1A_2}=\lim_{n\to 1}I_{A_1A_2}^{(n)}.
\end{equation}
It is found that the mutual information will mix the quantum and classical information together \cite{Plenio2007}, and hence
is not a good entanglement measure for mixed states.

Another quantity under extensive study, which is useful in characterizing the quantum entanglement in mixed states,
is the entanglement negativity
\cite{Vidal2002,Plenio2005}.
To be concrete, for a reduced density matrix
$\rho_{A_1A_2}$ which describes a mixed state in the Hilbert space $\mathcal{H}_{A_1}\otimes \mathcal{H}_{A_2}$,
a partial transposition of $\rho_{A_1A_2}$ with respect to the degrees of freedom in region $A_2$ is defined as
\begin{equation}\label{partialT}
\langle e_i^{(1)}e_j^{(2)}|\rho_{A_1\cup A_2}^{T_2}|e_k^{(1)}e_l^{(2)}\rangle=
\langle e_i^{(1)}e_l^{(2)}|\rho_{A_1\cup A_2}|e_k^{(1)}e_j^{(2)}\rangle,
\end{equation}
where $T_2$ represents the partial transposition over $A_2$, $|e_i^{(1)}\rangle$ and $|e_j^{(2)}\rangle$ are arbitrary bases in $\mathcal{H}_{A_1}$
and $\mathcal{H}_{A_2}$, respectively. 
Then the entanglement negativity is defined as
\begin{equation}\label{D1}
\mathcal{E}_{A_1A_2}:=\ln\text{tr}\left|\rho_{A_1\cup A_2}^{T_2}\right|.
\end{equation}
To calculate the entanglement negativity in a quantum filed theory, 
it is convenient to use the replica trick as follows \cite{Calabrese_CFT1,Calabrese_CFT2}
\begin{equation}\label{D2}
\mathcal{E}_{A_1A_2}=\lim_{n_e\to 1}\ln \text{tr}\left(\rho_{A_1\cup A_2}^{T_2}\right)^{n_e},
\end{equation}
where $n_e$ is an even integer.

Recently, the entanglement negativity has been extensively studied in conformal field theories \cite{Calabrese_CFT1,Calabrese_CFT2,Calabrese_finiteT}, quantum spin chain systems \cite{spinEN,spinENb},
coupled harmonic oscillators in one and two dimensions \cite{HarmonicO,HarmonicO2d,HarmonicO2db},
free fermion systems \cite{Eisler,Coser1508,Herzog,Chang},
topological ordered systems \cite{Vidal_2013, Catelnovo_2013,WenCS}, and holographic entanglement \cite{Holography,
Holography1,Holography2}.
Furthermore, the entanglement negativity has also been studied in the non-equilibrium case \cite{Eisler1406, Coser1410,Hoogeveen1412,Wen1501}
as well as the finite temperature case \cite{Eisler1406,Calabrese1408,Sherman1510}.

In this work, we focus on the topological entanglement negativity in
a particular topological quantum field theory (TQFT)--the Chern-Simons theory \cite{Witten1989,Witten1992}.
TQFTs are extensively used in condensed matter physics because of the emergence
of topological phases from many-body systems such as
the fractional quantum Hall states \cite{FQH}, gapped quantum spin
liquids \cite{QSL}, $p_x + ip_y$ superconductors \cite{Pwave} and so on.
It is noted that the entanglement negativity for a toric code model, an Abelian topological ordered system, has been studied
in previous works \cite{Vidal_2013, Catelnovo_2013}.
Later, an edge theory approach was developed
to study the entanglement negativity (and other entanglement measures) in a Chern-Simons theory.
Due to the bulk-boundary correspondence in a Chern-Simons theory,
we present an alternative approach to study
the entanglement negativity from bulk point of view.
By applying the replica trick and the surgery method \cite{Witten1989,Dong},
we compute the entanglement negativity in a generic
Chern-Simons theory.

The rest of the paper is organized as follows.
In Sec.\ \ref{Path Integral}, we introduce the path integral representation
of a partially transposed reduced density matrix, which is used to define the entanglement negativity.
In Sec.\ \ref{CS}, we introduce the basic ingredients of Chern-Simons theory and the surgery method.
Then by using the surgery method and the replica trick,
 we calculate the entanglement negativity on various bi-/tri-partitioned manifolds for a
Chern-Simons theory in Sec.\ref{EN calculation},
and study how the entanglement negativity depends on the
presence of quasiparticles and the choice of ground states.
Then we conclude in Sec.\ \ref{conclusion}.
We also include several appendices containing the calculation
of the entanglement negativity for a bipartitioned torus,
which is helpful to understand the case of a tripartite torus in the main text.
To compare with the entanglement negativity,
we also calculate the mutual information for various cases in the
appendices.

\subsection{Path integral representation of partially transposed reduced density matrix}
\label{Path Integral}

In this section, we introduce the path integral representation of a partially transposed reduced density matrix,
\textit{i.e.},
$\rho_{A_1\cup A_2}^{T_2}$ [see Eq.\ (\ref{partialT})],
as well as $ \text{tr}\left(\rho_{A_1\cup A_2}^{T_2}\right)^{n_e}$
[see Eq.\ (\ref{D2})].
The definition in this part applies to a generic quantum field theory.

The density matrix in a thermal state can be expressed as a path integral in the imaginary time interval $(0, \beta)$  \cite{Calabrese2004,Calabrese_CFT2}
\begin{align}
\rho\Big[\{\varphi_0(\vec{x})\},\{\varphi_{\beta}(\vec{x})\}\Big]
&=\frac{1}{Z(\beta)}
\left\langle
\{\varphi_0(\vec{x})\}|e^{-\beta H}|\{\varphi_{\beta}(\vec{x})\}\right\rangle
\nonumber \\
&=\int\prod [d\phi (\vec{x},\tau)]e^{-S_E}\prod_{\vec{x}}\delta[\phi(\vec{x},0)-\varphi_0(\vec{x})]
 \delta[\phi(\vec{x},\beta)-\varphi_{\beta}(\vec{x})],
\end{align}
with
$\beta\to\infty$ corresponding to the zero temperature limit.
Here $S_E$ is the Euclidean action and $Z=\text{tr}e^{-\beta H}$ is the partition function.
$\vec{x}$ represents the
coordinate in the $d$-dimensional space, and $\tau$ is the imaginary time.
The rows and columns of the density matrix are represented by the values
of fields $\phi(\vec{x},\tau)$ at $\tau=0$ and $\beta$, respectively.
Now we take partial transposition corresponding to a subsystem $B$,
then the partial transposed density matrix may be expressed as
\begin{align}
\quad
\rho^{T_B}\Big[\{\varphi_0(\vec{x})\},\{\varphi_{\beta}(\vec{x})\}\Big]
&=\int\prod_{\vec{x}, \tau} [d\phi (\vec{x},\tau)]e^{-S_E}
\prod_{\vec{x}\notin B}\delta[\phi(\vec{x},0)-\varphi_0(\vec{x})]\delta[\phi(\vec{x},\beta)-
\varphi_{\beta}(\vec{x})]
\nonumber \\
&\quad\times
\prod_{\vec{x}\in B}\delta[\phi(\vec{x},0)-\varphi_{\beta}(\vec{x})]\delta[\phi(\vec{x},\beta)-
\varphi_{0}(\vec{x})].
\end{align}
Suppose the system is tripartitioned into $A_1$, $A_2$ and $B$, then the
reduced density matrix for $\rho_{A=A_1\cup A_2}$ can be obtained by tracing $B$, \textit{i.e.},
\begin{align}
&\rho_{A_1\cup A_2}\left[\{\varphi_0(\vec{x})\}, \{\varphi_{\beta}(\vec{x})\}\Big| \vec{x}\in A_1\cup A_2
\right]
\nonumber \\
&=
\int \left(
\prod_{\vec{x}\in B}[d \varphi_0(\vec{x}) d\varphi_{\beta}(\vec{x})]\delta[\varphi_0(\vec{x})-\varphi_{\beta}(\vec{x})]
\right)
\rho\Big[\{\varphi_0(\vec{x})\},\{\varphi_{\beta}(\vec{x})\}\Big].
\end{align}
Then the partial transposed reduced density matrix $\rho_{A_1\cup A_2}^{T_2}$ over the subregion $A_2$ may be
expressed as
\begin{align}
&\quad
\rho_{A_1\cup A_2}^{T_{A_2}}\left[\{\varphi_0(\vec{x})\}, \{\varphi_{\beta}(\vec{x})\}\Big| \vec{x}\in A_1\cup A_2
\right]
\nonumber \\
&=
\int \left(
\prod_{\vec{x}\in B}[d \varphi_0(\vec{x}) d\varphi_{\beta}(\vec{x})]\delta[\varphi_0(\vec{x})-\varphi_{\beta}(\vec{x})]
\right)
\rho^{T_{A_2}}\Big[\{\varphi_0(\vec{x})\},\{\varphi_{\beta}(\vec{x})\}\Big].
\end{align}
Then, $\text{tr}\left(\rho_{A_1\cup A_2}^{T_{A_2}}\right)^n$ can be obtained by taking $n$ copies of $\rho_{A_1\cup A_2}^{T_{A_2}}$, and glue them appropriately as follows
\begin{align}
&\quad
\text{tr}\left(\rho_{A_1\cup A_2}^{T_{A_2}}\right)^n
=\int \prod_{k=1}^n\Bigg\{
\prod_{\vec{x}}\left[d\varphi_0^{(k)}(\vec{x})d\varphi_{\beta}^{(k)}(\vec{x})\right]
\prod_{\vec{x}\in B}\delta\left[
\varphi_0^{(k)}(\vec{x})-\varphi_{\beta}^{(k)}(\vec{x})
\right]
\nonumber \\
&\times \prod_{\vec{x}\in A_1}\delta\left[\varphi_0^{(k)}(\vec{x})-\varphi_{\beta}^{(k+1)}(\vec{x})\right]
\prod_{\vec{x}\in A_2}\delta\left[\varphi_{\beta}^{(k)}(\vec{x})-\varphi_{0}^{(k+1)}(\vec{x})\right]
\rho\left[\{\varphi_0^{(k)}(\vec{x})\},\{\varphi_{\beta}^{(k)}(\vec{x})\}\right]
\Bigg\},
\end{align}
with $i\equiv i$ mod $n$.
As a comparison, it should be noted that $\text{tr}\left(\rho_{A_1\cup A_2}\right)^n$ has the expression
\begin{equation}
\begin{split}
\text{tr}\Big(\rho_{A_1\cup A_2}\Big)^n
=&\int \prod_{k=1}^n\Bigg\{
\prod_{\vec{x}}\left[d\varphi_0^{(k)}(\vec{x})d\varphi_{\beta}^{(k)}(\vec{x})\right]
\prod_{\vec{x}\in B}\delta\left[
\varphi_0^{(k)}(\vec{x})-\varphi_{\beta}^{(k)}(\vec{x})
\right]\\
&\times \prod_{\vec{x}\in A}\delta\left[\varphi_0^{(k)}(\vec{x})-\varphi_{\beta}^{(k+1)}(\vec{x})\right]
\rho\left[\{\varphi_0^{(k)}(\vec{x})\},\{\varphi_{\beta}^{(k)}(\vec{x})\}\right]
\Bigg\}.
\end{split}
\end{equation}
Once we obtain $\text{tr}\left(\rho_{A_1\cup A_2}^{T_{A_2}}\right)^n$,
we can calculate the entanglement negativity based on Eq.\ (\ref{D2}).

\subsection{Chern-Simons theory and surgery}
\label{CS}

Here we mainly review the properties of Chern-Simons theory that will be used in our study of the entanglement negativity.
One may refer to the seminal paper \cite{Witten1989} for details of the Chern-Simons theory.
The Chern-Simons theory action with a gauge group $G$ on a three-manifold $M$ is given by
\begin{equation}
S_{\text{CS}}=\frac{k}{4\pi}\int_{M}\text{tr}\left(
A\wedge dA+\frac{2}{3}A\wedge A\wedge A
\right),
\end{equation}
where `$\text{tr}$' is the trace over the fundamental representation of the gauge group $G$, $A$ is the $G$-connection on
a genetic three-manifold $M$,
and $k$ is the coupling constant, which is quantized.
The Chern-Simons theory is a {\it topological field theory} in the sense that
the correlation functions do not depend on the metric of the manifold $M$.
Since the action of the Chern-Simons theory does not contain the metric, the
partition function
\begin{equation}
Z(M)=\int [\mathcal{D} A] \mathrm{e}^{i S_{\text{CS}}(A)},
\end{equation}
can define a topological invariant of the manifold $M$.
Besides the partition function as an invariant of three-manifolds,
invariants of links and knots in three-manifolds can be also defined
in the Chern-Simons theory.
Such a link or a knot in three-manifolds
are the ``Wilson line", that traces the holonomy of the gauge connection
on an oriented closed curve $\mathcal{C}$ in a given irreducible representation $\hat{R}$
of $G$,
\begin{align}
W_R^{\mathcal{C}} (A)= \mathrm{tr}_{R} P \exp \int_{\mathcal{C}}A.
\end{align}
We can compute the correlation functions of non-intersecting links/knots $\mathcal{C}_i$,
$i=1,\cdots, N$, with a representation $\hat{R}_i$ to each $\mathcal{C}_i$ on a three-manifold $M$,
\begin{align}
Z(M, \hat{R}_1, \cdots, \hat{R}_N)
= \langle W_{\hat{R}_1}^{\mathcal{C}_1} \cdots  W_{\hat{R}_N}^{\mathcal{C}_N} \rangle
= \int [\mathcal{D} A ] \left( \prod_{i=1}^{N}  W_{\hat{R}_i}^{\mathcal{C}_i} \right)  \mathrm{e}^{i S_{\text{CS}}}.
\end{align}
(When necessary, 
we denote the partition function as
$Z(M, [\hat{R}_1, \cdots, \hat{R}_N]_{\mathcal{C}_1,\cdots,\mathcal{C}_N})$
where $[\hat{R}_1, \cdots, \hat{R}_N]_{\mathcal{C}_1,\cdots,\mathcal{C}_N}$ indicates the configuration of the links/knots of Wilson loops.)
These links/knots correlation functions can be seen as the partition functions of a Chern-Simons theory
on a three-manifold $M$ in the presence of Wilson loops.
As shown by Witten \cite{Witten1989}, the the partition functions are exactly calculable by
canonical quantization and the surgery.


The key ingredient of computing the partition function is
canonical quantization of a Chern-Simons theory
on a three-manifold $M$ with boundary given by a Riemann surface $\Sigma$.
This canonical quantization will produce a Hilbert space $\mathcal{H}_{\Sigma}$
with an associated state $|\Psi_M \rangle$.
The dual Hilbert space $\mathcal{H}^*_{\Sigma}$ with an associated state  $\langle \Psi_M|$ state can be obtained by reversing the orientation of the
$\Sigma$.
The partition function of a Chern-Simons theory on a (closed) three-manifold can be computed by
performing the Heegaard splitting, which
decomposes the three-manifold as the connected sum of two three-manifolds $M_1$
and $M_2$ with common boundary $\Sigma$.
The original three-manifold $M=M_1 \bigcup_f M_2$ is obtained by gluing
$M_1$ and $M_2$ through their boundary under the homeomorphism $f:\Sigma \to \Sigma$.
This homeomorphism acting in the Hilbert space can be presented by an operator
$U_f: \mathcal{H}_\Sigma \to  \mathcal{H}_\Sigma$.
Hence the partition can be evaluated as
\begin{align}
Z(M)= \langle \Psi_{M_2} | U_f| \Psi_{M_1} \rangle.
\end{align}
When the boundary is a sphere, \textit{i.e.}, $\Sigma=S^2$, the Hilbert space $\mathcal{H}_{S^2}$ is one dimensional.
When the boundary $\Sigma=T^2$, which can be seen as the boundary
of a solid torus $\mathbf{T}=D \times S^1$,
one can obtain a state in $\mathcal{H}_{T^2}$ by inserting a Wilson loop in the representation $\hat{R}_i$ around the non-contractible cycle in the solid torus,
\begin{align}
|\Psi_{\mathbf{T}, \hat{R}_i}\rangle=|\hat{R}_i \rangle.
\end{align}
The state without the Wilson loop is the vacuum state, denoted as $|\hat{0} \rangle$.

The above results allow us to compute the partition function on three-manifolds
in the presence of Wilson loops.
Let us start with $S^2 \times S^1$,
which can be seen as gluing two solid tori ${\bf T}=D \times S^1$ with boundaries identified.
I.e., $S^2$ comes from gluing two discs together along their boundary $S^1$.
The partition function of a Chern-Simons theory in this three-manifold is
\begin{align}
Z(S^2 \times S^1) = \langle \hat{0}|\hat{0} \rangle=1.
\end{align}

Performing the modular transformation $S$: $\tau\to -\frac{1}{\tau}$ on the second solid torus,
where $\tau$ is the modular parameter of the torus,
and gluing it back,
\textit{i.e.}, the non-contractible cycle of the first solid torus  is
homologous to the contractible cycle of the second solid torus,
we get $S^3$.
We obtain the Chern-Simons partition function
\begin{align}
Z(S^3) = \langle \hat{0}|S|\hat{0} \rangle=\mathcal{S}_{00},
\end{align}
where $\mathcal{S}_{ij}$ is the element of the modular $\mathcal{S}$ matrix.
If there is a Wilson loop in the representation $\hat{R}_i$ in one solid torus,
the Chern-Simons partition functions become
\begin{align}
&Z(S^2 \times S^1,\hat{R}_i)= \langle \hat{0}|\hat{R}_i \rangle = \delta_{0,i},
\notag\\
&Z(S^3,\hat{R}_i)= \langle \hat{0}|S|\hat{R}_i \rangle = \mathcal{S}_{0i}.
\end{align}

One can also consider a Wilson loop in the representation $\hat{R}_i$ in a solid torus,
which is glued to another solid torus with a Wilson loop in the representation $\hat{R}_j$.
The Chern-Simons partition functions are
\begin{align}
&Z(S^2 \times S^1,\hat{R}_i,\hat{R}_j)= \langle \hat{R}_i |\hat{R}_j \rangle = \delta_{i,j}.  \notag\\
&Z(S^3,\hat{R}_i,\hat{R}_j)= \langle \hat{R}_i|S|\hat{R}_j \rangle = \mathcal{S}_{ij}.
\end{align}

\begin{figure}[ttt]
   \begin{center}
     \includegraphics[height=4cm]{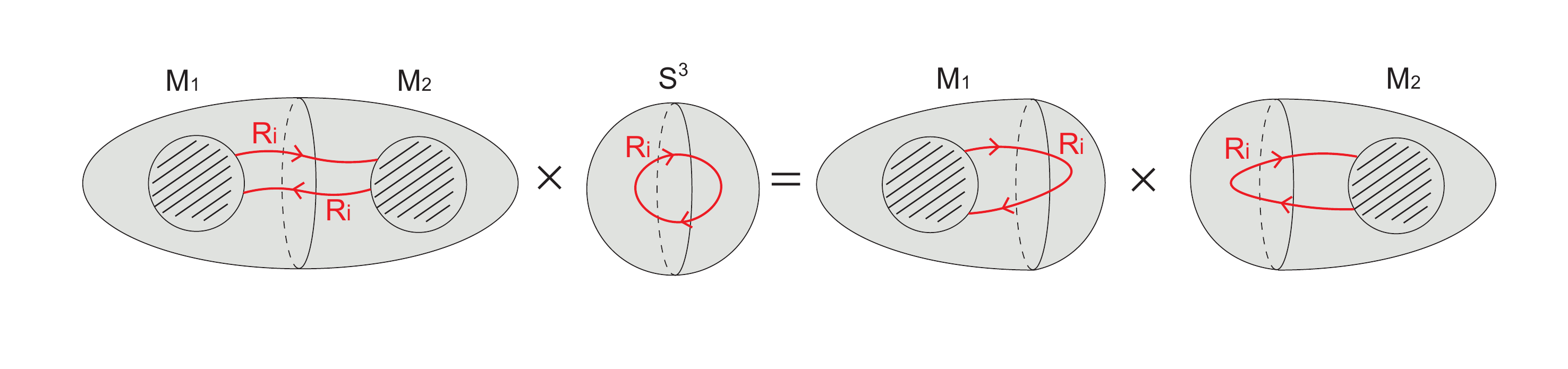}
   \end{center}
   \caption{The surgery procedure to relate the partition function on a manifold $M_1\cup M_2$ with the partition functions
on $M_1$ and $M_2$. 
The shaded region in $M_{1(2)}$ contains Wilson loops  with
a general links/knots configuration $\mathcal{C}_{1(2)}$.
}\label{surgeryW}
\end{figure}

Here we list two main properties of the above results:
\begin{enumerate}

\item The normalized vacuum expectation values of disjointed Wilson loops can be {\it factorized}, \textit{i.e.},
\begin{align}\label{nWilsonLine}
\frac{Z(M, \hat{R}_1, \cdots,  \hat{R}_N)}{Z(S^3)}=\prod_{i=1}^{N} \frac{Z(M_i, \hat{R}_i)}{Z(S^3)},
\end{align}
where the three-manifold $M$ is the connected sum of $N$
three-manifolds $M_i$ joined along $N-1$ two spheres $S^2$.
This result comes from the fact that the Hilbert space for $S^2$ is one-dimensional.

\item If Wilson loops are linked or they are passing through the common boundary $S^2$ between $M_i$
and $M_j$, the {\it factorizability} of the partition function is hold when
the Hilbert space for $S^2$ with a pair of charges in the dual representations $\hat{R}_i$ and $\hat{\overline{R}}_i$ is one-dimensional. We have
\begin{align}
Z(M,[\blacksquare_1,\blacksquare_2,\hat{R}_i, \hat{\overline{R}}_i]_{\mathcal{C}}) \cdot Z(S^3,\hat{R}_i) &= Z(M_1,[\blacksquare_1,\hat{R}_i]_{\mathcal{C}_1})
 \cdot Z(M_2,[\blacksquare_2,\hat{R}_i]_{\mathcal{C}_2}),
 \label{Eq: Z}
\end{align}
where $\blacksquare_{1(2)}$ contains Wilson loops with
a general links/knots configuration $\mathcal{C}_{1(2)}$ in the shaded region in the $M_{1(2)}$ manifold
shown in Fig.\ \ref{surgeryW}.
\end{enumerate}

A surgery procedure we will frequently use in this work is Eq.\ (\ref{Eq: Z}). We relate the partition
function on a manifold $M_1\cup M_2$ with the partition functions on $M_1$ and $M_2$, by a factor $Z(S^3,\hat{R}_i)=\mathcal{S}_{0i}$.
Notice that we do not consider any links/knots configuration of Wilson loops in our following discussion.
This indicates in our discussion, 
$\blacksquare_{1(2)}$ in Eq.\ (\ref{Eq: Z}) only contains unlinked/unknoted Wilson loops.

In addition, the modular $\mathcal{S}$-matrix, which is unitary, is related with the \textit{quantum dimension} as follows
\begin{align}
d_a=\frac{\mathcal{S}_{0a}}{\mathcal{S}_{00}}.
\end{align}
The unitarity condition for the $\mathcal{S}$-matrix implies that
\begin{align}
\left(\mathcal{S}_{00}\right)^{-1}=\sqrt{\sum_i|d_i|^2}=:\mathcal{D}.
\end{align}

\section{Topological entanglement negativity}
\label{EN calculation}

Based on the above discussion,
we study the topological entanglement negativity between two spatial subregions
on various manifolds in this section.
The entanglement negativity is calculated in the following steps.
(1) We consider $\text{tr}\left(\rho_{A_1\cup A_2}^{T_2}\right)^{n_e}$ as the partition function on a three-manifold $M$.
(2) We use the surgery method to compute $\text{tr}\left(\rho_{A_1\cup A_2}^{T_2}\right)^{n_e}$, and then take $n_e \to 1$.

To avoid confusions, a spatial manifold is a two-manifold, which can be viewed as the boundary
of the three-dimensional spacetime manifold where the wave function is defined.

\subsection{Bipartition of a sphere}

\begin{figure}[ttt]
   \begin{center}
     \includegraphics[height=6cm,natwidth=610,natheight=642]{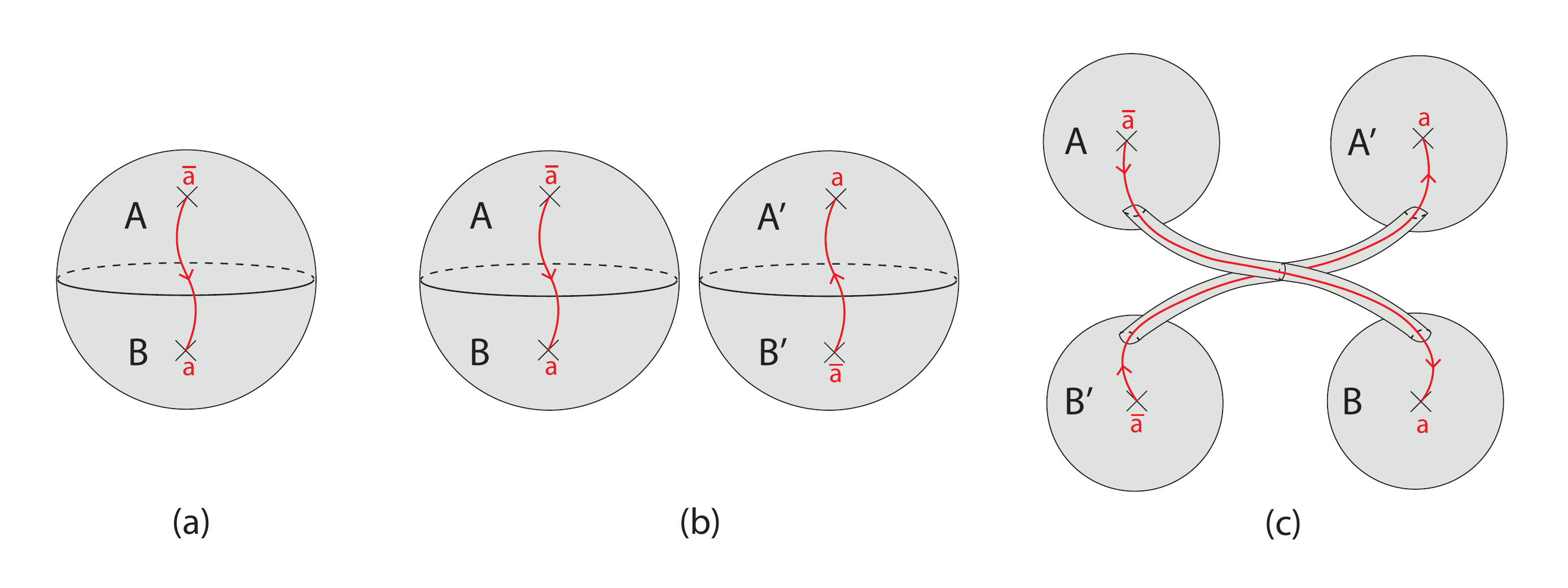}
   \end{center}
   \caption{(a)  Wave functional $|\Psi\rangle$. A Wilson line in representation $\hat{R}_a$ threads through the $AB$ interface. Shading implies a three-ball.  (b) $\rho_{A\cup B}=|\Psi\rangle\langle\Psi|$.  (c) $\rho_{A\cup B}^{T_{B}}$, in which we take partial transpose over $B$, \textit{i.e.}, we switch $B$ with $B'$.
}\label{sphere1}
\end{figure}

In this part, for the pedagogical purpose, we consider the simplest case, in which the spatial manifold is a two-sphere $S^2$.
We consider the general case where there is a quasiparticle $\bar{a}$ ($a$) in the subsystem $A$ ($B$),
where $\bar{a}$ is the anti-quasiparticle of $a$, \textit{i.e.}, $a\times \bar{a}=I+\cdots$, with $I$ being the identity operator.
A Wilson line in the representation $\hat{R}_a$ connects the quasiparticles $\bar{a}$ and $a$ at the two ends,
as shown in Fig.\ \ref{sphere1} (a).
For the case without quasiparticles, we can simply set $\bar{a}=a=I$ at the end.

Fig.\ \ref{sphere1} (a) represents the wave functional $|\Psi\rangle$, which is defined on a three-ball.
 It should be noted that the Wilson line in the representation
$\hat{R}_a$ is inside the solid ball.
For the density matrix $\rho=|\Psi\rangle\langle \Psi|$, we simply need to consider one more
3-ball with two conjugate punctures, which represents $\langle \Psi|$,
as shown in Fig.\ \ref{sphere1} (b).
To study the topological entanglement negativity between $A$ and $B$, we need to consider the partially transposed density matrix $\rho^{T_B}$ (or $\rho^{T_A}$).
Pictorially,  this can be operated by switching the submanifold $B$ and $B'$ as shown
in Fig.\ \ref{sphere1} (c).
Similar graphic representations of $\rho^{T_B}$ were also used in the tensor network study
of the entanglement negativity.
\cite{spinENb}

Next, to calculate the entanglement negativity between $A$ and $B$,
we will use the replica trick [see Eq.\ (\ref{D2})].
$\text{tr}\left(\rho^{T_B}\right)^{n}$ can be calculated as follows.
First, we  make $n$ copies of $\rho^{T_B}$,
with each copy represented in Fig.\ \ref{sphere1} (c).
Next, we glue the subregion $A'$ ($B$)
in the $i$-th copy with the subregion $A$ ($B'$) in the $(i+1)$-th (mod $n$) copy,
then we obtain $\text{tr}\left(\rho^{T_B}\right)^n$.
It is emphasized that $\text{tr}\left(\rho^{T_B}\right)^n$ depends on whether $n$ is odd or even.
For odd $n$,\textit{ i.e.}, $n=n_o$, the manifold after gluing is a $S^3$. On the other hand, for even $n$, \textit{i.e.},
$n=n_e$, the manifold
after gluing are two independent $S^3$. Therefore, $\text{tr}\left(\rho^{T_B}\right)^n$ after the normalization has the following
expressions
\begin{align}
\frac{\text{tr}\left(\rho^{T_{B}}\right)^{n_o}}
{\left(\text{tr}\rho^{T_{B}}\right)^{n_o}}
=&\frac{Z(S^3,\hat{R}_a)}{Z(S^3,\hat{R}_a)^{n_o}}
=Z(S^3,\hat{R}_a)^{1-n_o}=\left(\mathcal{S}_{0a}\right)^{1-n_o},
\nonumber \\
\frac{\text{tr}\left(\rho^{T_{B}}\right)^{n_e}}
{\left(\text{tr}\rho^{T_{B}}\right)^{n_e}}
=&\frac{Z(S^3,\hat{R}_a)^2}{Z(S^3,\hat{R}_a)^{n_e}}
=Z(S^3,\hat{R}_a)^{2-n_e}=\left(\mathcal{S}_{0a}\right)^{2-n_e},
\end{align}
where we have considered the fact that $\text{tr}\left(\rho^{T_B}\right)=Z(S^3,\hat{R}_a)=\mathcal{S}_{0a}$.
Then, according to the definition in Eq.\ (\ref{D2}), one can obtain the entanglement negativity as follows
\begin{equation}\label{ENsphere1}
\mathcal{E}_{AB}=\lim_{n_e\to 1}\ln \frac{\text{tr}\left(\rho^{T_{B}}\right)^{n_e}}
{\left(\text{tr}\rho^{T_{B}}\right)^{n_e}}=\ln \mathcal{S}_{0a}=\ln d_a-\ln \mathcal{D}.
\end{equation}
For the case without any quasiparticles on the sphere, one simply sets $d_a=d_I=1$, and therefore
\begin{equation}\label{ENsphere2}
\mathcal{E}_{AB}=-\ln \mathcal{D}.
\end{equation}
As a comparison, for odd $n$, one will obtain the trivial result, \textit{i.e.}, $\lim_{n_o\to 1}\ln \frac{\text{tr}\left(\rho^{T_{B}}\right)^{n_o}}
{\left(\text{tr}\rho^{T_{B}}\right)^{n_o}}=0$. It is noted that $\mathcal{E}_{AB}$ in Eqs.(\ref{ENsphere1})
and (\ref{ENsphere2}) are the same
as the topological entanglement entropy. This is because for a general pure state, the entanglement negativity for a bipartite
system is equal to the $1/2$ Renyi entropy, $\mathcal{E}_{AB}=S^{(1/2)}_A=S^{(1/2)}_B$. It is known that for the
case in Fig.\ \ref{sphere1} (a), one has $S^{(n)}_A=S^{(n)}_B=\ln d_a-\ln \mathcal{D}$ for arbitrary $n$.

Here we demonstrate the simplest case of computing the entanglement negativity by the surgery method.
As will be shown later, this basic operation provides a building block for the study of more complicated cases.

\subsection{Tripartition of a sphere}

\begin{figure}[ttt]
   \begin{center}
     \includegraphics[height=7cm]{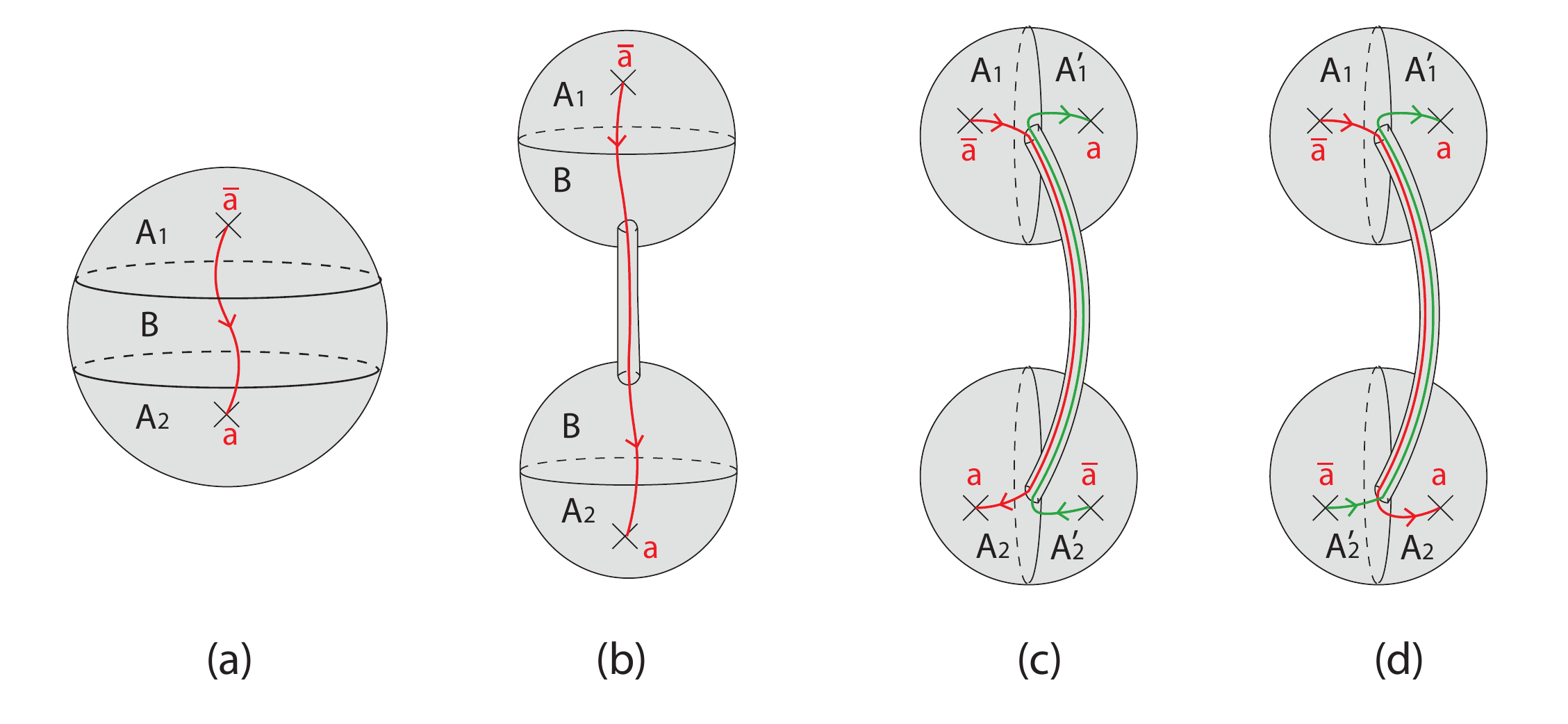}
   \end{center}
   \caption{(a)  Wave functional $|\Psi\rangle$.
A Wilson line in representation $\hat{R}_a$ threads through the  interface $A_1B$ and $A_2B$.
$A_1$ and $A_2$ are disjoint.
 (b) A three-manifold which is topologically equivalent to (a). (c) $\rho_{A_1\cup A_2}=\text{tr}_B|\Psi\rangle\langle\Psi|$, and
(d) $\rho_{A_1\cup A_2}^{T_{A_2}}$, in which we switch region $A_2$ and $A_2'$ in (c).
}\label{spheretriDisjoint}
\end{figure}

In this section, we study the entanglement negativity between $A_1$ and $A_2$ for
 a tripartite spatial manifold $S^2$, where the sphere is divided into $A_1$, $A_2$ and $B$.
In particular, we are mainly interested in two cases:  (1) $A_1$ and $A_2$ are adjacent, as shown in
Fig.\ \ref{spheretri1} (a), and (2) $A_1$ and $A_2$ are disjoint, as shown in Fig.\ \ref{spheretriDisjoint} (a).

\subsubsection{Case of adjacent $A_1$ and $A_2$}

\begin{figure}[ttt]
   \begin{center}
     \includegraphics[height=7cm]{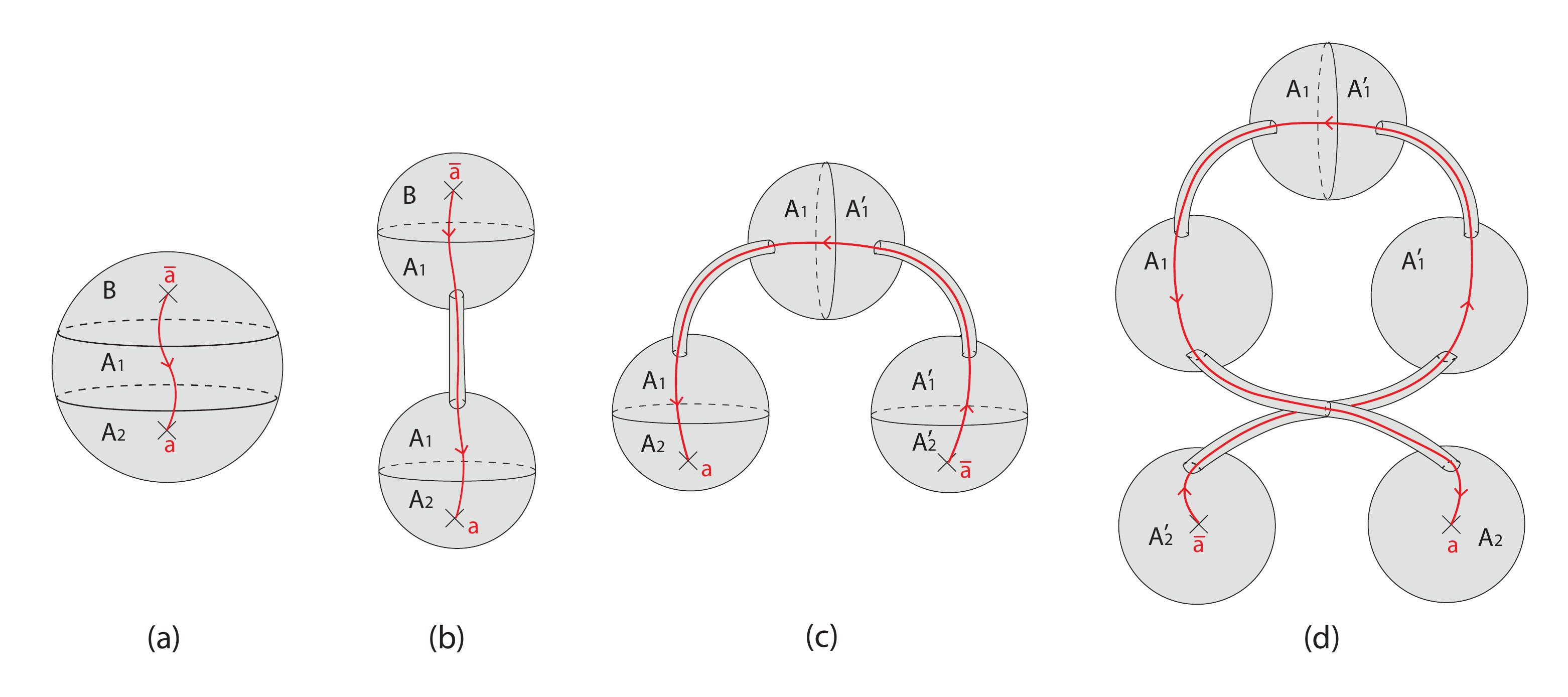}
   \end{center}
   \caption{(a)  Wave functional $|\Psi\rangle$. A Wilson line in representation $\hat{R}_a$ threads through the  interface $A_1B$ and $A_2B$, respectively. (b) A three-manifold which is topologically equivalent to (a). (c) $\rho_{A_1\cup A_2}=\text{tr}_B|\Psi\rangle\langle\Psi|$, and
(d) $\rho_{A_1\cup A_2}^{T_{A_2}}$, in which we do partial transposition over $A_2$, \textit{i.e.}, we switch
$A_2$ with $A_2'$ in (c).}\label{spheretri1}
\end{figure}

First we consider the case where $A_1$ and $A_2$ are adjacent to each other, as shown in Fig.\ \ref{spheretri1} (a).
There is a Wilson line in representation $\hat{R}_a$ which threads through both the $A_1A_2$ interface and the $A_1B$ interface.  Again, for the case without any quasiparticle on the sphere, one can simply set $\hat{R}_a=\hat{R}_I$ at the end.

For convenience, we deform the three-dimensional spacetime manifold in Fig.\ \ref{spheretri1} (a),
without changing the topology,
to two three-balls connected by a tube, as shown in
Fig.\ \ref{spheretri1} (b).
Then the reduced density matrix $\rho_{A_1\cup A_2}$ can be obtained by tracing over the $B$ part, as shown in Fig.\ \ref{spheretri1} (c).
Based on $\rho_{A_1\cup A_2}$, one can easily obtain $\rho_{A_1\cup A_2}^{T_{A_2}}$
by switching $A_2$ and $A_2'$, as shown in  Fig.\ \ref{spheretri1} (d).
One can find that the operation of partial transposition here is the same as that in Fig.\ \ref{sphere1}.

Now we are ready to calculate $\text{tr}\left(\rho_{A_1\cup A_2}^{T_{A_2}}\right)^n$ as follows.
We make $n$ copies of $\rho_{A_1\cup A_2}^{T_{A_2}}$, and glue the region $A_1'$ ($A_2$) in the $i$-th copy with $A_1$
($A_2'$) in the $(i+1)$-th (mod $n$) copy.
Similar with the case of a bipartitioned sphere,
the result depends on whether $n$ is odd or even as follows.

For odd $n$,\textit{ i.e.}, $n=n_o$, the resulting manifold is two $S^3$ connected by $n_o$ tubes. Each tube is contributed by
the one that connects $A_1'$ and $A_1'$ in Fig.\ \ref{spheretri1} (d). Then, by using the surgery procedure in Fig.\ \ref{surgeryW},
we cut all the tubes that connect the two $S^3$, with each tube contributing a factor $Z(S^3,\hat{R}_a)^{-1}$. Therefore, one can obtain
\begin{equation}
\begin{split}
\frac{\text{tr}\left(\rho_{A_1\cup A_2}^{T_{A_2}}\right)^{n_o}}
{\left(\text{tr}\rho_{A_1\cup A_2}^{T_{A_2}}\right)^{n_o}}
=&\frac{1}{Z(S^3,\hat{R}_a)^{n_o}}\cdot \frac{Z(S^3,\hat{R}_a)^2}{Z(S^3,\hat{R}_a)^{n_o}}
=Z(S^3,\hat{R}_a)^{2-2n_o}=\left(\mathcal{S}_{0a}\right)^{2-2n_o}.
\end{split}
\end{equation}
For even $n$, \textit{i.e.}, $n=n_e$,
the resulting manifold is \textit{three} $S^3$ connected by $n_e$ tubes.
The extra $S^3$ has the same origin as the case of a bipartitiond sphere in Fig.\ \ref{sphere1}.
Similar with the $n=n_o$ case,
by cutting each tube, one can obtain
\begin{equation}
\begin{split}
\frac{\text{tr}\left(\rho_{A_1\cup A_2}^{T_{A_2}}\right)^{n_e}}
{\left(\text{tr}\rho_{A_1\cup A_2}^{T_{A_2}}\right)^{n_e}}
=&\frac{1}{Z(S^3,\hat{R}_a)^{n_e}}\cdot \frac{Z(S^3,\hat{R}_a)^3}{Z(S^3,\hat{R}_a)^{n_e}}
=Z(S^3,\hat{R}_a)^{3-2n_e}=\left(\mathcal{S}_{0a}\right)^{3-2n_e}.
\end{split}
\end{equation}
Then the entanglement negativity between $A_1$ and $A_2$ can be expressed as
\begin{equation}\label{ENsphereTri1}
\mathcal{E}_{A_1A_2}=\lim_{n_e\to 1}\ln \frac{\text{tr}\left(\rho^{T_{B}}\right)^{n_e}}
{\left(\text{tr}\rho^{T_{B}}\right)^{n_e}}=\ln \mathcal{S}_{0a}=\ln d_a-\ln \mathcal{D},
\end{equation}
which is the same as Eq.\ (\ref{ENsphere1}).
In other words, for a tripartitioned $S^2$ as shown in Fig.\ \ref{spheretri1}, the
existence of region $B$ does not affect the entanglement negativity between $A_1$ and $A_2$, \textit{i.e.},
\begin{equation}\label{sphereEN0}
\mathcal{E}_{A_1A_2}(B\neq \emptyset)=\mathcal{E}_{A_1A_2}(B= \emptyset).
\end{equation}

\subsubsection{Case of disjoint $A_1$ and $A_2$}

Here, we consider the case that $A_1$ and $A_2$ are disjoint, as shown in Fig.\ \ref{spheretriDisjoint} (a).
We also include a quasiparticle $a$ (anti-quasiparticle $\bar{a}$)
in region $A_2$ ($A_1$). Therefore,
a Wilson line in representation $\hat{R}_a$ threads through both the $A_1B$ interface
and the $A_2B$ interface.

The three-manifold in  Fig.\ \ref{spheretriDisjoint} (a) is equivalent to two three-balls connected by a tube, as shown in
 Fig.\ \ref{spheretriDisjoint} (b). Based on this, one can obtain the reduced density matrix $\rho_{A_1\cup A_2}$,
as shown in Fig.\ref{spheretriDisjoint} (c). To get the partially transposed reduced density matrix $\rho_{A_1\cup A_2}^{T_{A_2}}$, one simply needs to switch $A_2$ with $A_2'$, as shown in  Fig.\ \ref{spheretriDisjoint} (d).

We now calculate the entanglement negativity between $A_1$ and $A_2$. As before, we make $n$ copies of
$\rho_{A_1\cup A_2}^{T_{A_2}}$ in Fig.\ \ref{spheretriDisjoint} (d).
Then we glue region $A_1'$ ($A_2$) in the $i$-th copy with the
region $A_1$ ($A_2'$) in the $(i+1)$-th (mod $n$) copy.
In this way, we obtain $\text{tr}\left(\rho_{A_1\cup A_2}^{T_{A_2}}\right)^n$. It can be found that the resulting manifold is two $S^3$ connected by $n$ tubes, which is independent
of whether $n$ is even or odd.
By considering the surgery procedure in Fig.\ \ref{surgeryW},
 one can cut
all the tubes that connect the two $S^3$, with each tube contributing a factor $Z(S^3,\hat{R}_a)$. Then one can obtain
\begin{equation}
\begin{split}
\frac{\text{tr}\left(\rho_{A_1\cup A_2}^{T_{A_2}}\right)^{n}}
{\left(\text{tr}\rho_{A_1\cup A_2}^{T_{A_2}}\right)^{n}}
=&\frac{1}{Z(S^3,\hat{R}_a)^{n}}\cdot \frac{Z(S^3,\hat{R}_a)^2}{Z(S^3,\hat{R}_a)^{n}}
=Z(S^3,\hat{R}_a)^{2-2n}=\left(\mathcal{S}_{0a}\right)^{2-2n},
\end{split}
\end{equation}
for both $n=n_o$ and $n=n_e$. Therefore, one can obtain the entanglement negativity between $A_1$ and $A_2$ as follows
\begin{equation}\label{ENsphereTri1}
\mathcal{E}_{A_1A_2}=\lim_{n_e\to 1}\ln \frac{\text{tr}\left(\rho^{T_{B}}\right)^{n_e}}
{\left(\text{tr}\rho^{T_{B}}\right)^{n_e}}=\ln \left(\mathcal{S}_{0a}\right)^0=0.
\end{equation}
I.e., there is no entanglement negativity between $A_1$ and $A_2$ in this case.
It is noted that the topological mutual information between $A_1$ and $A_2$ for this case is also zero [see Eq.\ (\ref{AppendixMIdisjoint})].

\subsection{Two adjacent non-contractible regions on a torus with non-contractible $B$}

\begin{figure}[ttt]
   \begin{center}
     \includegraphics[height=7.5cm]{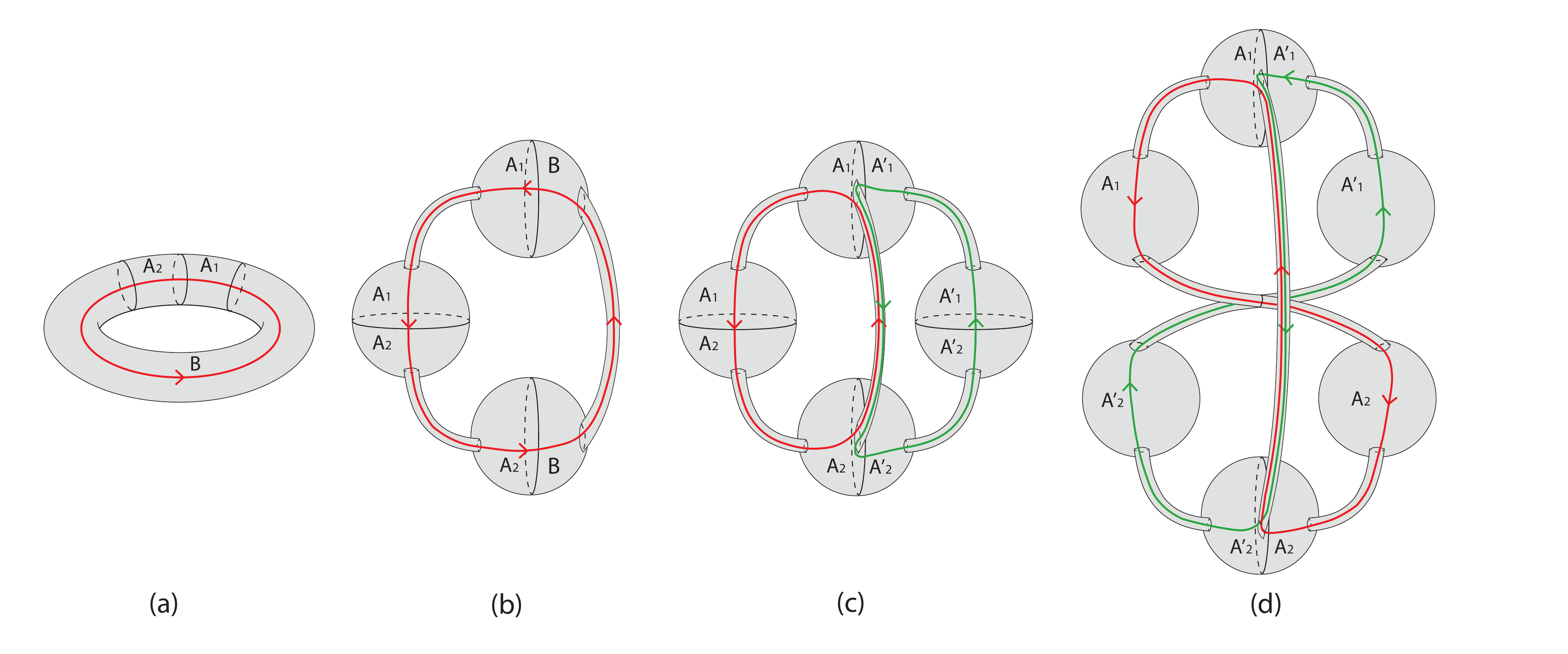}
   \end{center}
   \caption{(a)  Wave functional $|\Psi\rangle$. The toroidal space is divided into threes parts $A_1$, $A_2$ and $B$, where we have a one-component
$A_1A_2$ interface.
The red solid line represents a Wilson loop which can fluctuate among different representations.
(b) A three-manifold with three $3$-balls joined by three tubes appropriately, which is topologically equivalent to (a). (c) $\rho_{A_1\cup A_2}=\text{tr}_B|\Psi\rangle\langle\Psi|$, and
(d) $\rho_{A_1\cup A_2}^{T_{A_2}}$, in which we switch $A_2$ with $A_2'$ in (c).
}\label{torus3a}
\end{figure}

Here, we focus on the spatial manifold of a torus, $T^2$.
For the simplest case of a bipartite torus, one can refer to the Appendix A, where the operation is straightforward and helpful for
understanding the more complicated cases.

We first consider two adjacent non-contractible regions $A_1$ and $A_2$ on a torus with a non-contractible region $B$,
but with different number of components for the interface,  as shown  in Fig.\ \ref{torus3a} (a) and Fig.\ \ref{torus2c} (a).
In Fig.\ \ref{torus3a} (a), the two adjacent regions $A_1$ and $A_2$ share a one-component $A_1A_2$ interface,
and in Fig.\ \ref{torus2c} (a), the two adjacent regions $A_1$ and $A_2$ share a two-component $A_1A_2$ interface.
In the following, we will study the entanglement negativity between $A_1$ and $A_2$ for these two cases separately.

\subsubsection{One-component interface}

For the configuration in Fig.\ \ref{torus3a} (a), it is equivalent to three 3-balls
connected by three tubes, as shown in Fig.\ \ref{torus3a} (b).
Then one can obtain the reduced density matrix $\rho_{A_1\cup A_2}$
by tracing over the $B$ part, as shown in Fig.\ \ref{torus3a} (c). The partial transposition of the
reduced density matrix $\rho_{A_1\cup A_2}$ is fulfilled by switching $A_2$ with $A_2'$,
as shown in Fig.\ \ref{torus3a} (d).

Generally, the Wilson loop can fluctuate among different representations.
For simplicity, we first consider the case in which the Wilson loop is in a definite representation $\hat{R}_a$.
To study $\text{tr}\left(\rho_{A_1\cup A_2}^{T_{A_2}}\right)^{n}$, we make $n$ copies of $\rho_{A_1\cup A_2}^{T_{A_2}}$
in Fig.\ \ref{torus3a} (d). Then we glue region $A_1'$ ($A_2$) in the $i$-th copy with
$A_1$ ($A_2'$) in the $(i+1)$-th (mod $n$) copy, based on which we obtain $\text{tr}\left(\rho_{A_1\cup A_2}^{T_{A_2}}\right)^{n}$.
The result after gluing depends on whether $n$ is odd or even as follows: For odd $n$, \textit{i.e.}, $n=n_o$, the resulting manifold
is three $S^3$ connected by $3n_o$ tubes. The $3n_o$ tubes are contributed by the ones connecting $A_1'-A_1'$, $A_1-A_1$,
$A_2-A_2$, $A_2'-A_2'$ and $B-B$, respectively. The tube connecting $B-B$ corresponds to the vertical tube
in Fig.\ \ref{torus3a} (d). Then, by using the surgery procedure in Fig.\ \ref{surgeryW}, one can cut all the $3n_o$ tubes, with each tube contributing a factor $Z(S^3,\hat{R}_a)$. Then one can obtain
\begin{equation}
\begin{split}
\frac{\text{tr}\left(\rho_{A_1\cup A_2}^{T_{A_2}}\right)^{n_o}}
{\left(\text{tr}\rho_{A_1\cup A_2}^{T_{A_2}}\right)^{n_o}}
=&\frac{1}{Z(S^2\times S^1, \hat{R}_a,\hat{\overline{R}}_a)^{n_o}}\cdot\frac{Z(S^3,\hat{R}_a)^3}{Z(S^3,\hat{R}_a)^{3n_o}}
=Z(S^3,\hat{R}_a)^{3-3n_o}=\left(\mathcal{S}_{0a}\right)^{3-3n_o},
\end{split}
\end{equation}
where we have used the fact $Z(S^2\times S^1,\hat{R}_a,\hat{\overline{R}}_a)=1$. On the other hand, for even $n$, \textit{i.e.},
$n=n_e$, the resulting manifold is \textit{four} $S^3$ connected by $3n_e$ tubes, where the extra $S^3$ is caused by
the partial transposition. Similar with the case of $n=n_o$, the $3n_e$
tubes are contributed by the ones connecting $A_1'-A_1'$, $A_1-A_1$, $A_2-A_2$, $A_2'-A_2'$ and $B-B$, respectively.
By cutting all the $3n_e$ tubes with surgery, one can immediately obtain
\begin{equation}
\begin{split}
\frac{\text{tr}\left(\rho_{A_1\cup A_2}^{T_{A_2}}\right)^{n_e}}
{\left(\text{tr}\rho_{A_1\cup A_2}^{T_{A_2}}\right)^{n_e}}
=&\frac{1}{Z(S^2\times S^1, \hat{R}_a,\hat{\overline{R}}_a)^{n_e}}\cdot\frac{Z(S^3,\hat{R}_a)^4}{Z(S^3,\hat{R}_a)^{3n_e}}
=Z(S^3,\hat{R}_a)^{4-3n_e}=\left(\mathcal{S}_{0a}\right)^{4-3n_e}.
\end{split}
\end{equation}
It is then straightforward to show that for a general state $|\psi\rangle=\sum_j\psi_j|\hat{R}_j\rangle$, \textit{i.e.},
the Wilson loop is in a superposition of different representations $\hat{R}_j$,
one has
\begin{align}
\frac{\text{tr}\left(\rho_{A_1\cup A_2}^{T_{A_2}}\right)^{n_o}}
{\left(\text{tr}\rho_{A_1\cup A_2}^{T_{A_2}}\right)^{n_o}}
=&\frac{\sum_j|\psi_j|^{2n_o}\left(\mathcal{S}_{0j}\right)^{3-3n_o}}
{\left(\sum_j|\psi_j|^2\right)^{n_o}},
\nonumber \\
\frac{\text{tr}\left(\rho_{A_1\cup A_2}^{T_{A_2}}\right)^{n_e}}
{\left(\text{tr}\rho_{A_1\cup A_2}^{T_{A_2}}\right)^{n_e}}
=&\frac{\sum_j|\psi_j|^{2n_e}\left(\mathcal{S}_{0j}\right)^{4-3n_e}}
{\left(\sum_j|\psi_j|^2\right)^{n_e}}.
\end{align}
Then, one can obtain the entanglement negativity between $A_1$ and $A_2$ as follows
\begin{equation}
\mathcal{E}_{A_1 A_2}=\lim_{n_e\to 1}\ln \text{tr}\frac{\text{tr}\left(\rho_{A_1\cup A_2}^{T_{A_2}}\right)^{n_e}}
{\left(\text{tr}\rho_{A_1\cup A_2}^{T_{A_2}}\right)^{n_e}}=\ln\left(
\sum_j|\psi_j|^2\mathcal{S}_{0j}
\right)-\ln\sum_j|\psi_j|^2.
\end{equation}
By imposing the normalization condition $\sum_j|\psi_j|^2=1$, $\mathcal{E}_{A_1A_2}$ can be simplified as
\begin{equation}\label{ENtorus1component}
\mathcal{E}_{A_1 A_2}=\ln\left(
\sum_j|\psi_j|^2\mathcal{S}_{0j}
\right)=\ln \left(\sum_j|\psi_j|^2d_j\right)-\ln \mathcal{D}.
\end{equation}
Several comments on the above result are in orders:
\begin{enumerate}

\item 
By comparing with Eq.\ (\ref{ENbt}) for the case of a bipartitioned torus where $B=\emptyset$, it is found that
$\mathcal{E}_{A_1 A_2}(B\neq\emptyset)\neq\mathcal{E}_{A_1 A_2}(B=\emptyset)$,
which is different from the result for a tripartitioned sphere in Eq.\ (\ref{sphereEN0}).
The reason is that for a torus geometry, as the region $B$ shrinks to $\emptyset$, 
the component of the $A_1A_2$ interface changes from one to two.

\item
It is found that $\mathcal{E}_{A_1A_2}$ in Eq.\ (\ref{ENtorus1component}) can be used to distinguish an Abelian
theory from a non-Abelian theory. 
For an Abelian theory, we have $d_j=1$ for arbitrary representations $\hat{R}_j$,
and therefore
$\mathcal{E}_{A_1A_2}=-\ln \mathcal{D}$, which is independent of the choice of ground states. 
On the other hand,
for a non-Abelian theory, there exists at least one representation $\hat{R}_j$ so that $d_j>1$. 
Therefore, for a non-Abelian theory,
$\mathcal{E}_{A_1A_2}$ depends on the choice of ground states.

\end{enumerate}

\subsubsection{Two-component interface}

\begin{figure}[ttt]
   \begin{center}
     \includegraphics[height=10.6cm]{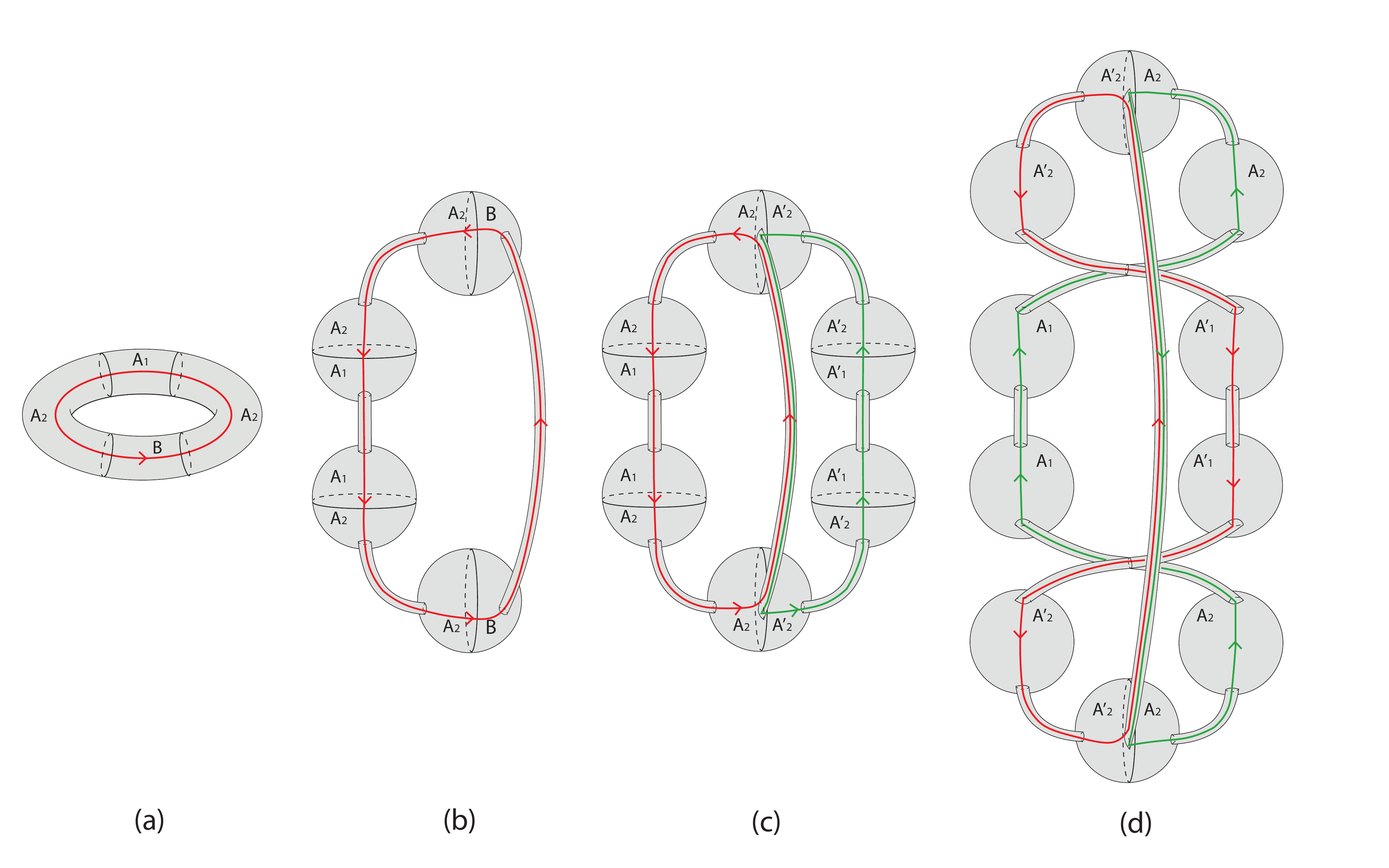}
   \end{center}
   \caption{(a) Wave functional $|\Psi\rangle$. 
   The toroidal space is divided into threes parts $A_1$, $A_2$ and $B$,  where we have a two-component
$A_1A_2$ interface.
The red solid line represents a Wilson loop which can fluctuate among different representations.
(b
) A three-manifold with four $3$-balls joined by four tubes appropriately, which is equivalent to the configuration
in (a) in topology. 
(c) $\rho_{A_1\cup A_2}=\text{tr}_B|\Psi\rangle\langle\Psi|$, and
(d) $\rho_{A_1\cup A_2}^{T_{A_2}}$, in which we do partial transposition over $A_2$,
\textit{i.e.}, we switch $A_2$ with $A_2'$ in (c).}\label{torus2c}
\end{figure}

Next, we consider two adjacent non-contractible regions $A_1$ and $A_2$ on a spatial manifold $T^2$,
with a two-component $A_1A_2$ interface, as shown in Fig.\ \ref{torus2c} (a).
The configuration in  Fig.\ \ref{torus2c} (a) is equivalent to four 3-balls connected by four tubes in topology,
as shown in  Fig.\ \ref{torus2c} (b). Then it is straightforward to obtain the reduced density
matrix $\rho_{A_1\cup A_2}$ by tracing out the $B$ part, as shown in  Fig.\ \ref{torus2c} (c). Next, for the partially transposed
reduced density matrix $\rho^{T_{A_2}}_{A_1\cup A_2}$, we simply need to switch $A_2$ with $A_2'$, as shown in
 Fig.\ \ref{torus2c} (d).

As in the previous part, we first consider the simple case that the Wilson
loop is in a definite representation $\hat{R}_a$.
To calculate $\text{tr}\left(\rho_{A_1\cup A_2}^{T_{A_2}}\right)^{n}$, we make $n$ copies of $\rho_{A_1\cup A_2}^{T_2}$ in Fig.\ \ref{torus2c} (d).
Then by gluing the region $A_1'$($A_2$) in the $i$-th copy with the region $A_1$($A_2'$) in the $(i+1)$-th (mod $n$) copy,
we can obtain  $\text{tr}\left(\rho_{A_1\cup A_2}^{T_{A_2}}\right)^{n}$.
As before,  the resulting manifold depends on whether $n$ is odd or even, as follows.
For odd $n$, \textit{i.e.}, $n=n_o$, the resulting manifold is four $S^3$ connected by $4n_o$ tubes.
 The $4n_o$ tubes are contributed by the ones connecting $A_2-A_2$, $A_1'-A_1'$, $A_1-A_1$,
$A_2'-A_2'$ and $B-B$, respectively. By using the surgery procedure in Fig.\ \ref{surgeryW},
we can cut all the $4n_o$ tubes,
with each tube contributing $Z(S^3,\hat{R}_a)$. Therefore, one can obtain
\begin{equation}
\begin{split}
\frac{\text{tr}\left(\rho_{A_1\cup A_2}^{T_{A_2}}\right)^{n_o}}
{\left(\text{tr}\rho_{A_1\cup A_2}^{T_{A_2}}\right)^{n_o}}
=&\frac{1}{Z(S^2\times S^1, \hat{R}_a,\hat{\overline{R}}_a)^{n_o}}\cdot\frac{Z(S^3,\hat{R}_a)^{4}}{Z(S^3,\hat{R}_a)^{4n_o}}
=Z(S^3,\hat{R}_a)^{4-4n_o}=\left(\mathcal{S}_{0a}\right)^{4-4n_o}.
\end{split}
\end{equation}
On the other hand,  for even $n$, \textit{i.e.},
$n=n_e$, the resulting manifold is \textit{six} $S^3$ connected by $3n_e$ tubes, where the extra two $S^3$ is caused by
the partial transposition. 
Similar with the case of $n=n_o$, the  $4n_e$ tubes are contributed by the ones connecting
$A_2-A_2$, $A_1'-A_1'$, $A_1-A_1$, $A_2'-A_2'$ and $B-B$, respectively.
With the surgery method, one can immediately obtain
\begin{equation}
\begin{split}
\frac{\text{tr}\left(\rho_{A_1\cup A_2}^{T_{A_2}}\right)^{n_e}}
{\left(\text{tr}\rho_{A_1\cup A_2}^{T_{A_2}}\right)^{n_e}}
=&\frac{1}{Z(S^2\times S^1, \hat{R}_a,\hat{\overline{R}}_a)^{n_e}}\cdot\frac{Z(S^3,\hat{R}_a)^{6}}{Z(S^3,\hat{R}_a)^{4n_e}}
=Z(S^3,\hat{R}_a)^{6-4n_e}=\left(\mathcal{S}_{0a}\right)^{6-4n_e}.
\end{split}
\end{equation}
It is then straightforward to show that for a general state $|\psi\rangle=\sum_j\psi_j|\hat{R}_j\rangle$, one has
\begin{align}
\frac{\text{tr}\left(\rho_{A_1\cup A_2}^{T_{A_2}}\right)^{n_o}}
{\left(\text{tr}\rho_{A_1\cup A_2}^{T_{A_2}}\right)^{n_o}}
&=\frac{\sum_j|\psi_j|^{2n_o}\left(\mathcal{S}_{0j}\right)^{4-4n_o}}
{\left(\sum_j|\psi_j|^2\right)^{n_o}},
\nonumber \\
\frac{\text{tr}\left(\rho_{A_1\cup A_2}^{T_{A_2}}\right)^{n_e}}
{\left(\text{tr}\rho_{A_1\cup A_2}^{T_{A_2}}\right)^{n_e}}
&=\frac{\sum_j|\psi_j|^{2n_e}\left(\mathcal{S}_{0j}\right)^{6-4n_e}}
{\left(\sum_j|\psi_j|^2\right)^{n_e}}.
\end{align}
Then, one can obtain the entanglement negativity between $A_1$ and $A_2$ as follows
\begin{equation}
\mathcal{E}_{A_1 A_2}=\lim_{n_e\to 1}\ln \text{tr}\frac{\text{tr}\left(\rho_{A_1\cup A_2}^{T_{A_2}}\right)^{n_e}}
{\left(\text{tr}\rho_{A_1\cup A_2}^{T_{A_2}}\right)^{n_e}}=\ln\left(
\sum_j|\psi_j|^2\mathcal{S}_{0j}^2
\right)-\ln\sum_j|\psi_j|^2.
\end{equation}
By imposing the normalization condition $\sum_j|\psi_j|^2=1$, $\mathcal{E}_{A_1A_2}$ can be simplified as
\begin{equation}\label{ENtorus2component1}
\mathcal{E}_{A_1 A_2}=\ln\left(
\sum_j|\psi_j|^2\mathcal{S}_{0j}^2
\right)=\ln \left(\sum_j|\psi_j|^2d_j^2\right)-2\ln \mathcal{D}.
\end{equation}
Compared with the case of one-component $A_1A_2$ interface in the previous part [see Eq.\ (\ref{ENtorus1component})],
here the power of $\mathcal{S}_{0j}$ is changed from $1$ to $2$, which is caused by changing the number
of components in $A_1A_2$ interface. In addition, similar with the result in Eq.\ (\ref{ENtorus1component}),
$\mathcal{E}_{A_1A_2}$ in Eq.\ (\ref{ENtorus2component1}) can also be used to distinguish an Abelian theory from
a non-Abelian theory by studying its dependence on the choice of ground states.

\subsection{Two adjacent non-contractible regions on a torus with contractible $B$}

\begin{figure}[ttt]
   \begin{center}
     \includegraphics[height=6.5cm]{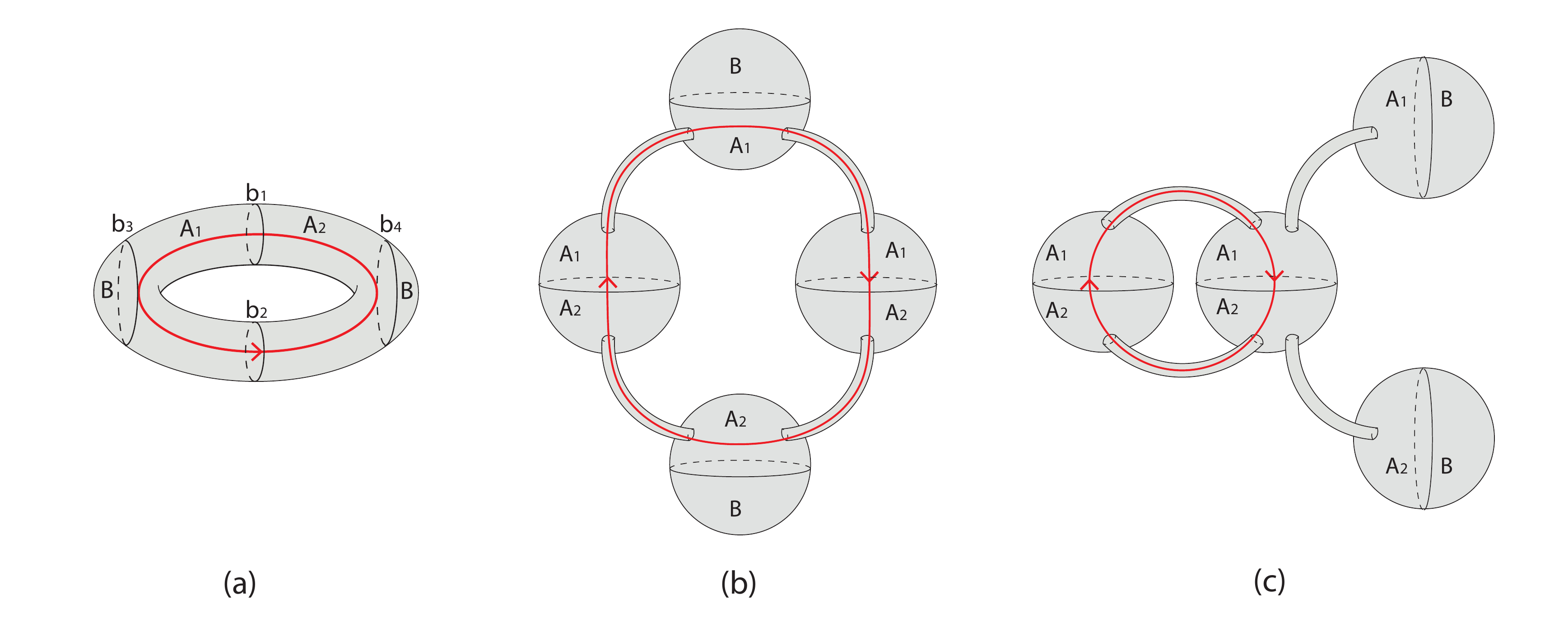}
   \end{center}
   \caption{(a)  Wave functional $|\Psi\rangle$. The toroidal space is divided into threes parts $A_1$, $A_2$ and $B$,  where we have a two-component
$A_1A_2$ interface and a contractible region $B$.
The red solid line represents a Wilson loop which can fluctuate among different representations.
(b) A three-manifold with four $3$-balls joined by four tubes appropriately, which is equivalent to the configuration
in (a) in topology. The configuration in (b) can be further deformed into the configuration in (c), without changing topology.
}\label{Torustri00}
\end{figure}

\begin{figure}[ttt]
   \begin{center}
     \includegraphics[height=9cm]{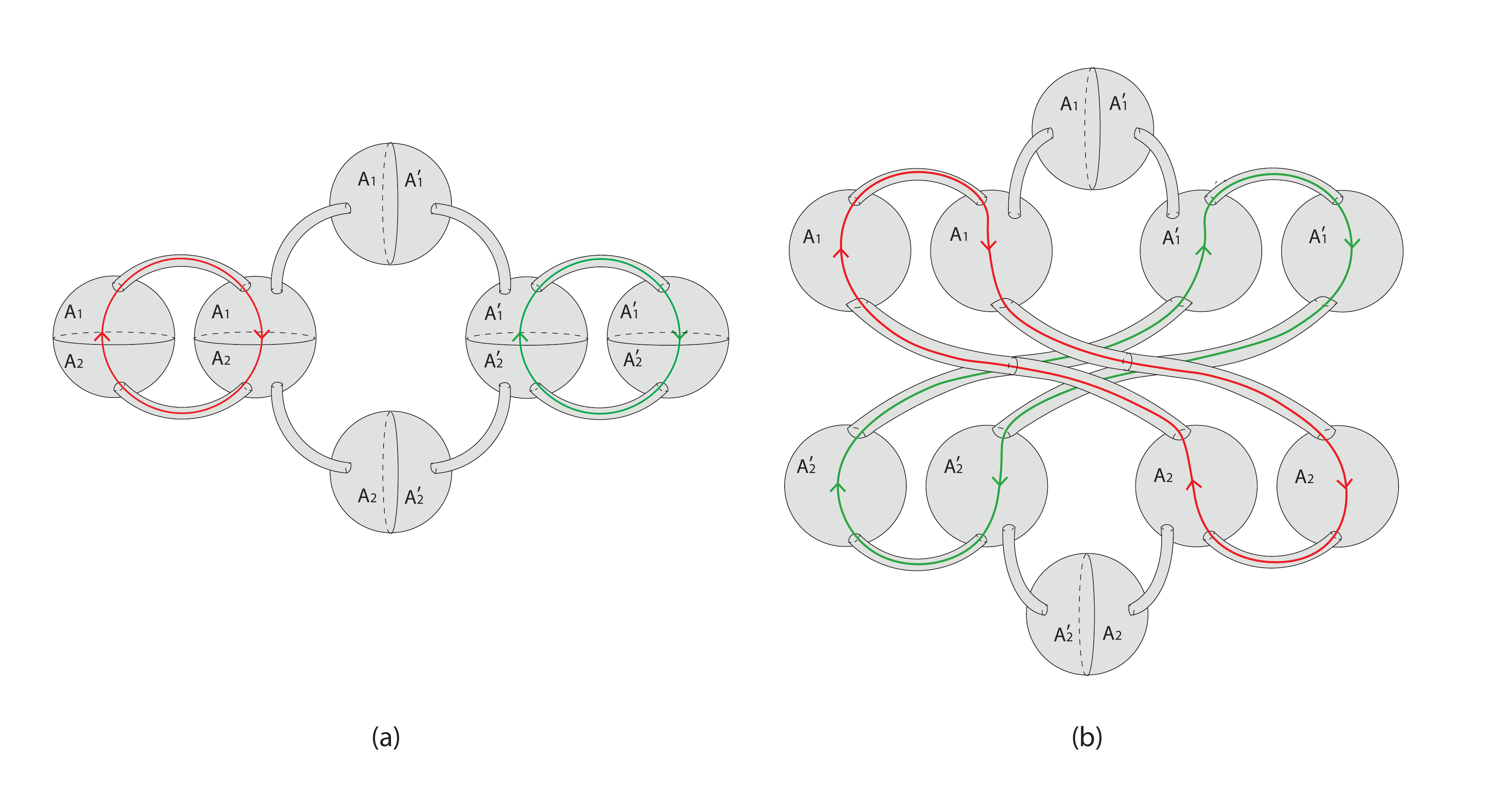}
   \end{center}
   \caption{(a) The reduced density matrix $\rho_{A_1\cup A_2}$, which is obtained based on the wave functional in Fig.\ref{Torustri00} (c).
 (b) The partially transposed reduced density matrix $\rho_{A_1\cup A_2}^{T_{A_2}}$, which is obtained
by switching $A_2$ and $A_2'$ in (a).}\label{Torustri00Density}
\end{figure}

Here, we study the entanglement negativity between two adjacent non-contractible regions $A_1$ and $A_2$
on a spatial manifold $T^2$, with a contractible region $B$, as shown in Fig.\ \ref{Torustri00} (a).
For convenience, we deform the three-manifold in  Fig.\ \ref{Torustri00} (a) into the three-manifold in
 Fig.\ \ref{Torustri00} (b), where there are four $S^3$ connected by four tubes, which can be further
deformed into the three-manifold in  Fig.\ \ref{Torustri00} (c). Then it is straightforward to obtain the
reduced density matrix $\rho_{A_1A_2}$ by tracing out the $B$ part, as shown in  Fig.\ \ref{Torustri00Density} (a).
To obtain the partially transposed reduced density matrix $\rho_{A_1\cup A_2}^{T_{A_2}}$, we simply need
to switch $A_2$ with $A_2'$ in $\rho_{A_1\cup A_2}$, as shown in Fig.\ \ref{Torustri00Density} (b).

As before, for simplicity, we first consider the case in which the Wilson loop is in a definite representation $\hat{R}_a$.
To obtain $\text{tr}\left(\rho_{A_1\cup A_2}^{T_{A_2}}\right)^n$, we make $n$ copies of $\rho_{A_1\cup A_2}^{T_{A_2}}$
in Fig.\ \ref{Torustri00Density} (b). Then we glue the region $A_1'$($A_2$) in the $i$-th copy with the region $A_1$($A_2'$) in the
$(i+1)$-th (mod $n$) copy, and obtain $\text{tr}\left(\rho_{A_1\cup A_2}^{T_{A_2}}\right)^n$.
Since the configuration in Fig.\ \ref{Torustri00Density} (b) is already very complicated,
it is helpful for the readers to understand the gluing based on
the case of a bipartite torus [see Fig.\ \ref{DisjointTorus01bb} (c)], considering that the limit $B\to \emptyset$
in Fig.\ \ref{Torustri00} (a) corresponds to a bipartitioned torus.

The gluing result depends on whether $n$ is odd or even as follows. For odd $n$, \textit{i.e.}, $n=n_o$, the resulting manifold is
four $S^3$ connected by $4n_o$ tubes. One should be very careful here.
For convenience, we label the four rows of 3-balls in Fig.\ \ref{Torustri00Density} (d) as the first, second, third and fourth rows of
3-balls from top to bottom.  In the resulting manifold after gluing, two $S^3$ are contributed by the
3-balls in the first and fourth rows  in  Fig.\ \ref{Torustri00Density} (d).
It is noted that there is no Wilson line threading through these two $S^3$, and therefore each of them contributes
$Z(S^3)$ after the surgery.
The other two $S^3$ are contributed by the
3-balls in the second and third rows. Since there are Wilson lines threading through these two $S^3$, each of them
contributes $Z(S^3,\hat{R}_a)$ after the surgery.

For the $4n_o$ tubes, $2n_o$ tubes are contributed by the ones that connect the first (third) and second (fourth) rows of 3-balls.
 There are no Wilson lines threading through
these $2n_o$ tubes. Therefore, after the surgery procedure in Fig.\ \ref{surgeryW}, each of these tubes
contributes $Z(S^3)$.
The other $2n_o$ tubes are contributed by the tubes that connect $A_1-A_1$ ($A_1'-A_1'$) in the second
row, and the ones that connect $A_2'-A_2'$ ($A_2-A_2$) in the third row.
For these $2n_o$ tubes, since there are Wilson lines threading through them, each tube contributes a factor
$Z(S^3,\hat{R}_a)$ after the surgery.

Based on the above analysis, one can obtain
\begin{align}\label{n0torus}
\frac{\text{tr}\left(\rho_{A_1\cup A_2}^{T_{A_2}}\right)^{n_o}}
{\left(\text{tr}\rho_{A_1\cup A_2}^{T_{A_2}}\right)^{n_o}}
&=\frac{1}{Z(S^2\times S^1; \hat{R}_a, \hat{\overline{R}}_a)^{n_o}}\cdot
\frac{Z(S^3)^2\cdot Z(S^3,\hat{R}_a)^2}{Z(S^3)^{2n_o}\cdot Z(S^3,\hat{R}_a)^{2n_o}}
\nonumber\\
&=Z(S^3)^{2-2n_o}\cdot Z(S^3,\hat{R}_a)^{2-2n_o}
\nonumber\\
&=\left(\mathcal{S}_{00}\right)^{2-2n_o}
\left(
\mathcal{S}_{0a}
\right)^{2-2n_o}.
\end{align}
On the other hand, for even $n$, \textit{ i.e.}, $n=n_e$, 
the resulting manifold is \textit{six} $S^3$ connected by $4n_e$ tubes.
Compared with the case of $n=n_o$, the extra two $S^3$ are introduced by the partial transposition, which is similar to
the case of a bipartite torus in Fig.\ \ref{DisjointTorus01bb}.
In particular, for the extra two $S^3$, there are Wilson lines
threading through them, and therefore each of them contributes $Z(S^3,\hat{R}_a)$ after the surgery. Therefore, one can obtain
\begin{align}
\frac{\text{tr}\left(\rho_{A_1\cup A_2}^{T_{A_2}}\right)^{n_e}}
{\left(\text{tr}\rho_{A_1\cup A_2}^{T_{A_2}}\right)^{n_e}}
&=\frac{1}{Z(S^2\times S^1; \hat{R}_a, \hat{\overline{R}}_a)^{n_e}}\cdot \frac{Z(S^3)^2}{Z(S^3)^{2n_e}}\cdot
\left[\frac{Z(S^3,\hat{R}_a)^2}{Z(S^3,\hat{R}_a)^{n_e}}\right]^2
\nonumber\\
&=Z(S^3)^{2-2n_e}\cdot Z(S^3,\hat{R}_a)^{4-2n_e}
\nonumber\\ 
&=\left(\mathcal{S}_{00}\right)^{2-2n_e}
\left(
\mathcal{S}_{0j}
\right)^{4-2n_e}.
\end{align}
It is emphasized that the square term $[\cdots]^2$ in the first row arises from the fact that
the two sets of Wilson loops (red and green) in Fig.\ \ref{Torustri00Density} (b), after gluing $2n_e$ copies, are
independent to each other. 
This square term is absent in Eq.\ (\ref{n0torus}), because the two sets of Wilson loops are glued to each other
for $n=n_o$.

It is straightforward to check that for a general state $|\psi\rangle=\sum_i\psi_i|\hat{R}_i\rangle$, one has
\begin{align}
\frac{\text{tr}\left(\rho_{A_1\cup A_2}^{T_{A_2}}\right)^{n_o}}
{\left(\text{tr}\rho_{A_1\cup A_2}^{T_{A_2}}\right)^{n_o}}
&=
\left(\mathcal{S}_{00}\right)^{2-2n_0}\cdot
\frac{\sum_j|\psi_j|^{2n_o}\left(\mathcal{S}_{0j}\right)^{2-2n_o}}
{\left(\sum_j|\psi_j|^2\right)^{n_o}},
\nonumber \\
\frac{\text{tr}\left(\rho_{A_1\cup A_2}^{T_{A_2}}\right)^{n_e}}
{\left(\text{tr}\rho_{A_1\cup A_2}^{T_{A_2}}\right)^{n_e}}
&=
\left(\mathcal{S}_{00}\right)^{2-2n_e}\cdot
\frac{\left[\sum_j|\psi_j|^{n_e}\left(\mathcal{S}_{0j}\right)^{2-n_e}\right]^2}
{\left(\sum_j|\psi_j|^2\right)^{n_e}}.
\label{Torus2even}
\end{align}
Then, one can obtain the entanglement negativity between $A_1$ and $A_2$ as follows
\begin{equation}
\mathcal{E}_{A_1 A_2}=\lim_{n_e\to 1}\ln \text{tr}\frac{\text{tr}\left(\rho_{A_1\cup A_2}^{T_{A_2}}\right)^{n_e}}
{\left(\text{tr}\rho_{A_1\cup A_2}^{T_{A_2}}\right)^{n_e}}=2\ln\left(
\sum_j|\psi_j|\mathcal{S}_{0j}
\right)-\ln\sum_j|\psi_j|^2.
\end{equation}
By imposing the normalization condition $\sum_j|\psi_j|^2=1$, $\mathcal{E}_{A_1A_2}$ can be simplified as
\begin{equation}\label{ENtorus2component}
\mathcal{E}_{A_1 A_2}=2\ln\left(
\sum_j|\psi_j|\mathcal{S}_{0j}
\right)=2\ln \left(\sum_j|\psi_j|d_j \right)-2\ln \mathcal{D}.
\end{equation}
The result is the same as Eq.\ (\ref{ENbt}) for a bipartite torus, \textit{i.e.}, $\mathcal{E}_{A_1A_2}(B\neq \emptyset)=
\mathcal{E}_{A_1A_2}(B=\emptyset)$ for the configuration in Fig.\ \ref{Torustri00} (a).
For this case, the entanglement negativity
between $A_1$ and $A_2$ depends on the choice of ground states for both Abelian and non-Abelian
Chern-Simons theories.

\begin{figure}[ttt]
   \begin{center}
     \includegraphics[height=9cm]{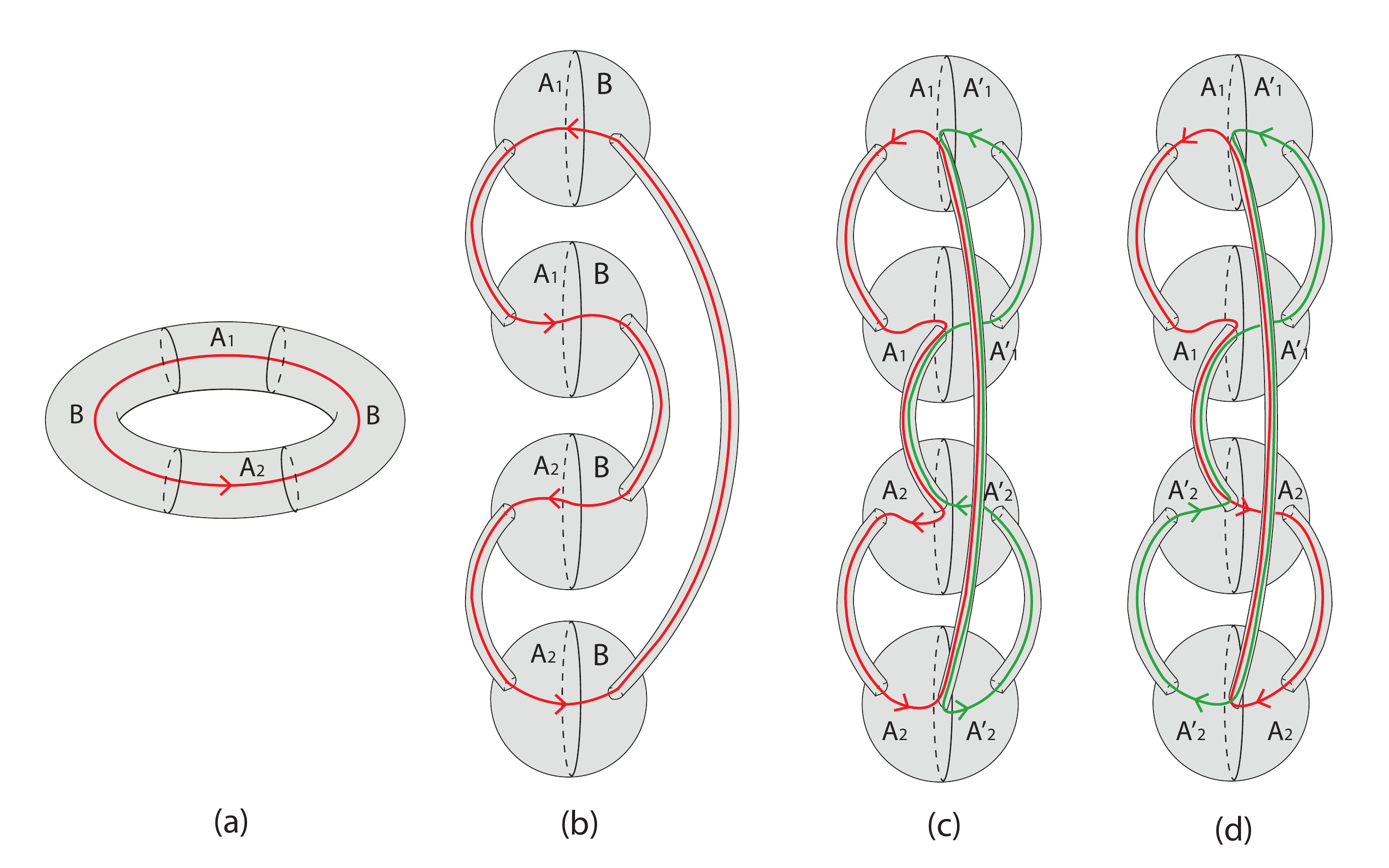}
   \end{center}
   \caption{(a) Wave functional $|\Psi\rangle$.
The toroidal space is divided into threes parts $A_1$, $A_2$ and $B$, where $A_1$ and $A_2$ are disjoint.
(b) The configuration in (a) is topologically equivalent to four $3$-balls joined by four tubes appropriately.
(c) $\rho_{A_1\cup A_2}$ by tracing out part $B$. (d) Partially transposed reduced density matrix
$\rho_{A_1\cup A_2}^{T_2}$ where the partial transposition is over degrees of freedom in $A_2$,
i.e., we switch $A_2$ with $A_2'$ in $\rho_{A_1\cup A_2}$ in (c).
}\label{DisjointTorus01}
\end{figure}

\subsection{Two disjoint non-contractible regions on a torus}

Finally, we demonstrate the vanishing entanglement negativity for two disjoint non-contractible regions $A_1$ and $A_2$ on a spatial manifold $T^2$, as shown in
Fig.\ \ref{DisjointTorus01} (a), in which the regions $A_1$ and $A_2$ are separated by non-contractible regions $B$.
The configuration in Fig.\ \ref{DisjointTorus01} (a) is topologically equivalent
to four $3$-balls connected by four tubes appropriately, as shown in Fig.\ \ref{DisjointTorus01} (b).
Then it is straightforward to obtain the reduced density matrix $\rho_{A_1\cup A_2}$ by tracing out the $B$ part,
as shown in Fig.\ \ref{DisjointTorus01} (c). The partially transposed reduced density matrix $\rho_{A_1\cup A_2}^{T_{A_2}}$
in Fig.\ \ref{DisjointTorus01} (d) is obtained by switching $A_2$ and $A_2'$ in Fig.\ \ref{DisjointTorus01} (c).

As before, we first consider the simple case that the Wilson loop is in a definite representation $\hat{R}_a$.
By repeating the gluing and surgery procedures as before,  it is straightforward to check that
\begin{align}
\frac{\text{tr}\left(\rho_{A_1\cup A_2}^{T_{A_2}}\right)^{n_o}}
{\left(\text{tr}\rho_{A_1\cup A_2}^{T_{A_2}}\right)^{n_o}}
&=\frac{1}{Z(S^2\times S^1; \hat{R}_a, \hat{\overline{R}}_a)^{n_o}}\cdot \frac{Z(S^3,\hat{R}_a)^4}{Z(S^3,\hat{R}_a)^{4n_o}}
=Z(S^3,\hat{R}_a)^{4-4n_o}=\left(
\mathcal{S}_{0a}
\right)^{4(1-n_o)},
\nonumber \\
\frac{\text{tr}\left(\rho_{A_1\cup A_2}^{T_{A_2}}\right)^{n_e}}
{\left(\text{tr}\rho_{A_1\cup A_2}^{T_{A_2}}\right)^{n_e}}
&=\frac{1}{Z(S^2\times S^1; \hat{R}_a, \hat{\overline{R}}_a)^{n_e}}\cdot \frac{Z(S^3,\hat{R}_a)^4}{Z(S^3,\hat{R}_a)^{4n_e}}
=Z(S^3,\hat{R}_a)^{4-4n_e}=\left(
\mathcal{S}_{0a}
\right)^{4(1-n_e)}.
\end{align}
The result is independent of whether $n$ is odd or even.
For a general state $|\psi\rangle=\sum_i\psi_i |\hat{R}_i\rangle$, one can find that
\begin{equation}
\begin{split}
\frac{\text{tr}\left(\rho_{A_1\cup A_2}^{T_{A_2}}\right)^{n_o(n_e)}}
{\left(\text{tr}\rho_{A_1\cup A_2}^{T_{A_2}}\right)^{n_o(n_e)}}
=
\frac{\sum_i |\psi_i|^{2n_o(n_e)}\left(
\mathcal{S}_{0i}
\right)^{4(1-n_o(n_e))}}{\left(\sum_i|\psi_i|^2\right)^{n_o(n_e)}}.
\end{split}
\end{equation}
Then, one can obtain the entanglement negativity between $A_1$ and $A_2$ as follows
\begin{equation}
\mathcal{E}_{A_1 A_2}=\lim_{n_e\to 1}\ln \frac{\text{tr}\left(\rho_{A_1\cup A_2}^{T_{A_2}}\right)^{n_e}}
{\left(\text{tr}\rho_{A_1\cup A_2}^{T_{A_2}}\right)^{n_e}}=0,
\end{equation}
\textit{i.e.}, there is no entanglement negativity between 
two disjoint non-contractible regions on a torus.

\section{Concluding remarks}
\label{conclusion}

In this work, by using the surgery method and the replica trick, 
we compute the topological entanglement negativity
between two spatial regions for Chern-Simons field theories. 
We study examples
on various manifolds with different bipartitions or tripartitions.
In particular, we study how the entanglement negativity depends on the distributions 
of quasiparticles and the choice of ground states.
For two adjacent non-contractible regions on a tripartitioned torus,
the entanglement negativity is dependent (independent) on the choice of ground states 
for non-Ablelian (Abelian) theories. 
Therefore, it provides a simple way to distinguish Abelian and non-Abelian theories.
Our method applies to arbitrary oriented (2+1) dimensional manifolds with arbitrary ways of bipartions/tripartitions.

All the cases studied in this work agree with 
the results obtained by using a complimentary approach, 
the edge theory approach, 
presented in Ref.\ \cite{WenCS}.
Here we would like to give some remarks on comparing
the method in this work and the edge theory approach:
(1) For the edge theory approach, it is unnecessary to know the (2+1) dimensional spacetime manifold and gluing pictures,
which are usually complicated. 
By expressing the (Ishibashi) edge states at the entanglement cut, a straightforward
calculation can be performed, which is usually tedious.
(2) On the other hand,  for the surgery approach shown  in this work, 
the only complication is understanding the 3-manifold.
However, this method is elegant, in the sense that once the corresponding
3-manifold for the partially transposed reduced density matrix is known, the results can be directly read off.
Thus, these two methods are complimentary and have their own merits.

Finally, we close by pointing out a future problem:
It is interesting to generalize our method(s) to higher dimensions, such as (3+1) dimensions.
Most recently, the surgery of (3+1)-dimensional manifolds
(with particle and loop excitations) 
was discussed in Ref.\ \cite{WangJ,WangJ2}. 
One can study the entanglement entropy and negativity of 
(3+1)-dimensional TQFTs, 
once the corresponding partition function can be evaluated.

\appendix

\section{Topological entanglement negativity: Bipartitioned torus}
\label{Bipartite torus}

\begin{figure}[ttt]
   \begin{center}
     \includegraphics[height=6cm]{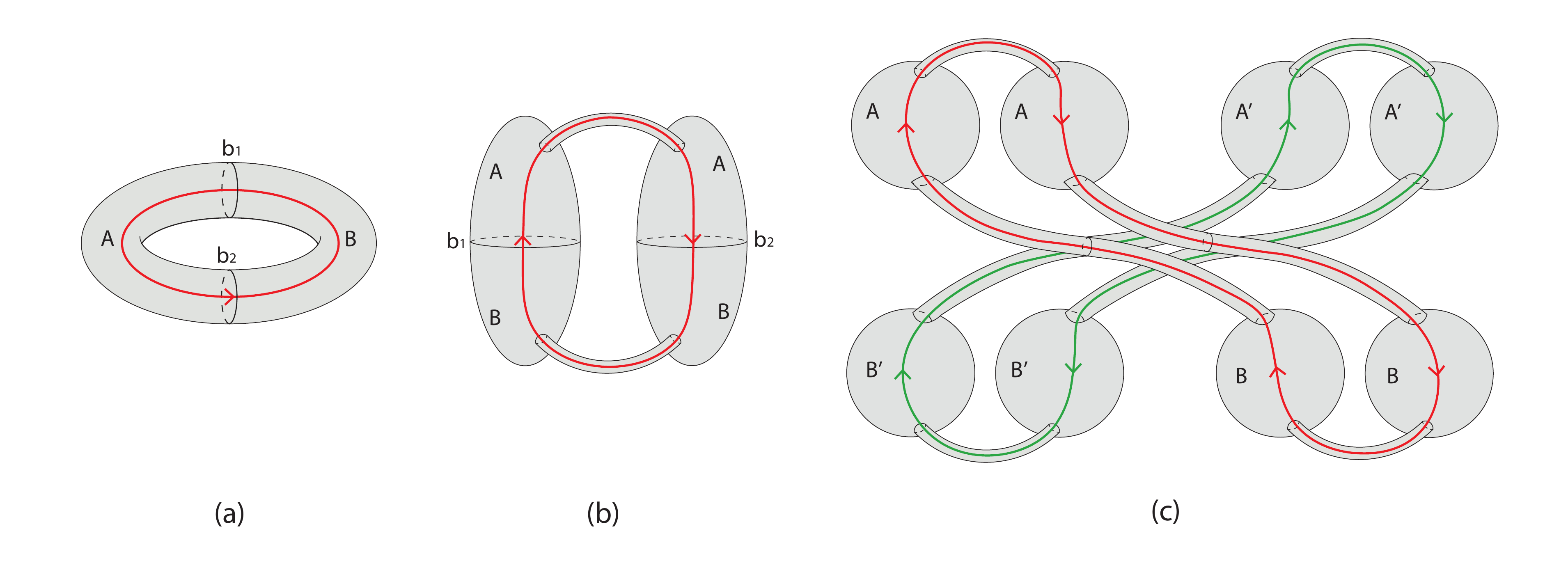}
   \end{center}
   \caption{
(a) Wave functional $|\Psi\rangle$. 
The toroidal space is bipartitioned into $A$ and $B$.
(b) The configuration in (a) can be deformed to two $3$-balls joined by two tubes appropriately.
(c) Partially transposed reduced density matrix
$\rho_{A\cup B}^{T_B}$, where the partial transposition is fulfilled by switching $B$ and $B'$.
}\label{DisjointTorus01bb}
\end{figure}

Here, we consider a bipartitioned torus, as shown in Fig.\ \ref{DisjointTorus01bb} (a), 
which is topologically
equivalent to two 3-balls connected by two tubes as shown in Fig.\ \ref{DisjointTorus01bb} (b).
It is straightforward to obtain the partially transposed reduced density matrix $\rho_{AB}^{T_B}$
as shown in Fig.\ \ref{DisjointTorus01bb} (c).

For the first step, we consider the simplest case, \textit{i.e.},
the Wilson loop is in a definite representation $\hat{R}_a$.
We follow the
replica trick introduced in the main text:
First, we make $n$ copies of $\rho_{AB}^{T_B}$  as in Fig.\ \ref{DisjointTorus01} (c). 
Then, we glue the region $A'$($B$) in the $i$-th copy 
with the region $A$($B'$) in the $(i+1)$-th (mod $n$) copy, 
based on which we obtain
$\text{tr}\left(\rho_{AB}^{T_B}\right)^n$. 
One can find that the resulting manifold depends on
whether $n$ is odd or even. 
For odd $n$, \textit{i.e.}, $n=n_o$, 
we obtain two $S^3$ connected by $2n_o$ tubes.
It should be noted that these tubes are contributed by those connecting $A-A$, $A'-A'$, $B-B$, and $B'-B'$ in Fig.\ \ref{DisjointTorus01bb} (c).
By considering the surgery procedure in Fig.\ \ref{surgeryW},
we cut all the $2n_o$ tubes, with each tube contributing $Z(S^3,\hat{R}_a)$. Then we can obtain
\begin{align}
\frac{\text{tr}\left(\rho_{A\cup B}^{T_{B}}\right)^{n_o}}
{\left(\text{tr}\rho_{A\cup B}^{T_{B}}\right)^{n_o}}
=&\frac{1}{Z(S^2\times S^1,\hat{R}_a,\hat{\overline{R}}_a)^{n_o}}\cdot
\frac{Z(S^3,\hat{R}_a)^2}{Z(S^3,\hat{R}_a)^{2 n_o}}
=Z(S^3,\hat{R}_a)^{2-2n_o}=\left(\mathcal{S}_{0a}\right)^{2-2n_o},
\end{align}
where we have used the fact that $\text{tr}\rho_{A\cup B}^{T_B}=Z(S^2\times S^1,\hat{R}_a,\hat{\overline{R}}_a)=1$.
On the other hand, for even $n$, \textit{i.e.}, $n=n_e$,
the resulting manifold is composed of two independent manifolds, with each
manifold being two $S^3$ connected by $n_e$ tubes.
By using the same surgery procedure as above, we can obtain
\begin{align}
\frac{\text{tr}\left(\rho_{A\cup B}^{T_{B}}\right)^{n_e}}
{\left(\text{tr}\rho_{A\cup B}^{T_{B}}\right)^{n_e}}
=&\frac{1}{Z(S^2\times S^1,\hat{R}_a,\hat{\overline{R}}_a)^{n_e}}\cdot\left[
\frac{Z(S^3,\hat{R}_a)^2}{Z(S^3,\hat{R}_a)^{n_e}}\right]^2
=Z(S^3,\hat{R}_a)^{4-2n_e}=\left(\mathcal{S}_{0a}\right)^{4-2n_e}.
\end{align}
It is then straightforward to show that for a general pure state
$|\psi\rangle=\sum_j\psi_j|\hat{R}_j\rangle$,
\textit{i.e.}, the Wilson loop is in a superposition of different representations, 
one has
\begin{align}
\frac{\text{tr}\left(\rho_{A\cup B}^{T_{B}}\right)^{n_o}}
{\left(\text{tr}\rho_{A\cup B}^{T_{B}}\right)^{n_o}}
=&\frac{\sum_j|\psi_j|^{2n_o}\left(\mathcal{S}_{0j}\right)^{2-2n_o}}
{\left(\sum_j|\psi_j|^2\right)^{n_o}},
\nonumber \\
\frac{\text{tr}\left(\rho_{A\cup B}^{T_{B}}\right)^{n_e}}
{\left(\text{tr}\rho_{A\cup B}^{T_{B}}\right)^{n_e}}
=&\frac{\left[\sum_j|\psi_j|^{n_e}\left(\mathcal{S}_{0j}\right)^{2-n_e}\right]^2}
{\left(\sum_j|\psi_j|^2\right)^{n_e}}.
\end{align}
Then one can obtain the entanglement negativity between $A$ and $B$ as follows
\begin{equation}
\mathcal{E}_{AB}=\lim_{n_e\to 1}\ln \text{tr}\frac{\text{tr}\left(\rho_{A\cup B}^{T_{B}}\right)^{n_e}}
{\left(\text{tr}\rho_{A\cup B}^{T_{B}}\right)^{n_e}}=2\ln\left(
\sum_j|\psi_j|\mathcal{S}_{0j}
\right)-\ln\sum_j|\psi_j|^2.
\end{equation}
By imposing the normalization condition $\sum_j|\psi_j|^2=1$, $\mathcal{E}_{AB}$ can
be simplified as
\begin{equation}\label{ENbt}
\mathcal{E}_{AB}=2\ln\left(
\sum_j|\psi_j|\mathcal{S}_{0j}
\right)
=2\ln\left(
\sum_j|\psi_j|d_j
\right)-2\ln \mathcal{D},
\end{equation}
which is the same as the $\frac{1}{2}$-Renyi entropy $S_A^{(1/2)}$ ($S_B^{(1/2)}$), as expected.

\section{Topological mutual information between two regions for various cases}
\label{MI}

Here, we make a comparison between the mutual information and 
the entanglement negativity between two regions $A_1$ and $A_2$ for various cases.
It is noted that the mutual information in Chern-Simons theories is also studied for some specific cases in 
a recent paper \cite{Jian}.

\subsection{Tripartitioned sphere}

\begin{figure}[ttt]
   \begin{center}
     \includegraphics[height=7cm]{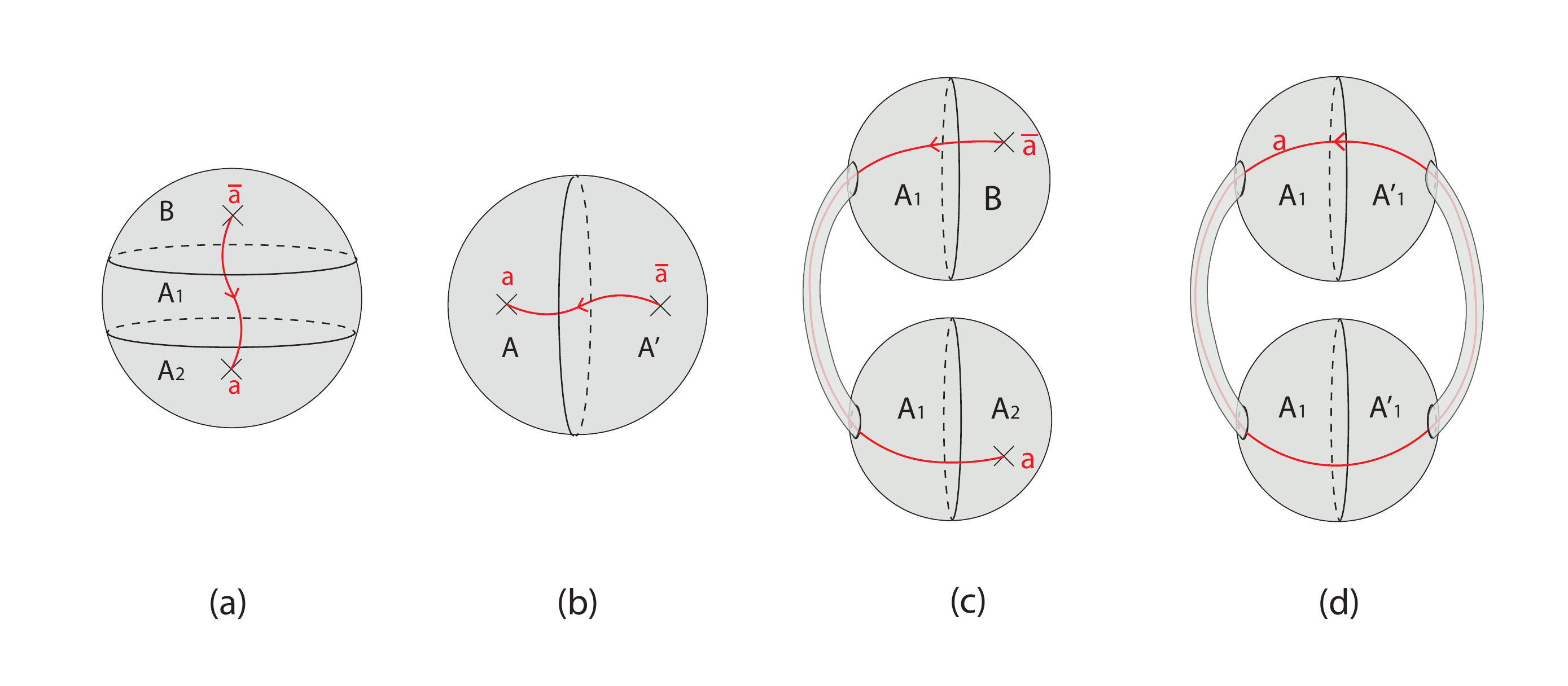}
   \end{center}
   \caption{
(a)  
Wave functional $|\Psi\rangle$. 
A Wilson line in representation $\hat{R}_a$ threads through the  interface $A_1B$ and $A_2B$, respectively.
(b) 
$\rho_A=\text{tr}_B|\Psi\rangle\langle\Psi|$. 
(c) 
Wave functional $|\Psi\rangle$, which is
topologically equivalent to (a). 
(d) $\rho_{A_1}=\text{tr}_{A_2\cup B}|\Psi\rangle\langle \Psi|$.
}\label{sphereTri1MIa}
\end{figure}

\subsubsection{Case of adjacent $A_1$ and $A_2$}

This case corresponds to the configuration in Fig.\ \ref{spheretri1} (a) 
[see also Fig.\ \ref{sphereTri1MIa} (a)].
First, we calculate the entanglement entropy for $A=A_1\cup A_2$. 
Based on the configuration of $\rho_A$
in Fig.\ \ref{sphereTri1MIa} (b), it is straightforward to check that
\begin{equation}
\text{tr}\left(\rho_A^n\right)=Z(S^3,\hat{R}_a)
\quad
\text{and}
\quad
\text{tr}\rho_A =Z(S^3,\hat{R}_a).
\end{equation}
Therefore,
\begin{equation}\label{SA003}
S_{A}^{(n)}=\frac{1}{1-n}\ln\frac{\text{tr}\left(\rho_A^n\right)}{\left(\text{tr}\rho_A\right)^n}
=\ln Z(S^3,R_a)=\ln \mathcal{S}_{0a}.
\end{equation}
Observing the topology in Fig.\ \ref{sphereTri1MIa} (a), it is straightforward to check that
$S_{A_2}^{(n)}=S_A^{(n)}$.
Next, to calculate the entanglement entropy for $A_1$, 
we deform the configuration of $|\Psi\rangle$ into
Fig.\ \ref{sphereTri1MIa} (c), 
based on which one can obtain $\rho_{A_1}$ in Fig.\ \ref{sphereTri1MIa} (d).
Then, one has
\begin{equation}
\text{tr}\left(\rho_{A_1}^n\right)=\frac{Z(S^3,\hat{R}_a)\cdot Z(S^3,\hat{R}_a)}{\left[Z(S^3,\hat{R}_a)\right]^n},
\end{equation}
and
\begin{equation}\label{SA1aa}
\begin{split}
S_{A_1}^{(n)}=&\frac{1}{1-n}\ln\frac{\text{tr}\left(\rho_{A_1}^n\right)}{\left(\text{tr}\rho_{A_1}\right)^n}\\
=&\frac{1}{1-n}\ln \frac{Z(S^3,\hat{R}_a)\cdot Z(S^3,\hat{R}_a)}{\left[Z(S^3,\hat{R}_a)\right]^{2n}}\\
=&2\ln\mathcal{S}_{0a}.
\end{split}
\end{equation}
Therefore, the mutual information between $A_1$ and $A_2$ is given by
\begin{equation}
I^{(n)}_{A_1,A_2}=S_{A_1}^{(n)}+S_{A_2}^{(n)}-S_{A_1\cup A_2}^{(n)}=2\ln\mathcal{S}_{0a}
=2\ln d_a-2\ln \mathcal{D},
\end{equation}
which is independent of $n$.

\subsubsection{Case of disjoint $A_1$ and $A_2$}

This case corresponds to the configuration in Fig.\ \ref{spheretriDisjoint},
and can be easily studied based on the previous results.
It can be found that $S_{A_1}^{(n)}$ and $S_{A_2}^{(n)}$ have the same form as $S_A^{(n)}$
in Eq.\ (\ref{SA003}).
In addition, since the total system stays in a pure state, then we have
$S_{A_1\cup A_2}^{(n)}=S_B^{(n)}$, which has
the same expression as Eq.\ (\ref{SA1aa}). 
Then one has
\begin{equation}\label{AppendixMIdisjoint}
I^{(n)}_{A_1,A_2}=S_{A_1}^{(n)}+S_{A_2}^{(n)}-S_{A_1\cup A_2}^{(n)}=0.
\end{equation}

\subsection{Two adjacent non-contractible regions on a torus with non-contractible $B$}

\subsubsection{One-component interface}

This case corresponds to the configuration in Fig.\ \ref{torus3a}(a), and
can be studied based on the result in Ref.\ \cite{Dong}.
For a general state $|\Psi\rangle=\sum_a\psi_a |\hat{R}_a\rangle$,
it is found that
\begin{align}\label{SA00a}
S_{A_1}^{(n)}&=S_{A_2}^{(n)}=S_{A_1\cup A_2}^{(n)}=\frac{1}{1-n}\ln \sum_a |\psi_a|^{2n} d_a^{2-2n}-2\ln \mathcal{D},\notag\\
\text{and}\quad
S_{A_1}&=S_{A_2}=S_{A_1\cup A_2}=2\sum_a|\psi_a|^2\ln d_a-\sum_a|\psi_a|^2\ln |\psi_a|^2-2\ln\mathcal{D}.
\end{align}
Then one can immediately obtain
\begin{align}
I^{(n)}_{A_1,A_2}&=\frac{1}{1-n}\ln \sum_a |\psi_a|^{2n} d_a^{2-2n}-2\ln \mathcal{D},\notag\\
 \quad\text{and}\quad
I_{A_1,A_2}&=2\sum_a|\psi_a|^2\ln d_a-\sum_a|\psi_a|^2\ln |\psi_a|^2-2\ln \mathcal{D}.
\end{align}

\subsubsection{Two-component interface}

This case corresponds to the configuration in Fig.\ \ref{torus2c} (a). 
$S^{(n)}_{A_1}$ and $S^{(n)}_{A_1\cup A_2}$ can be obtained from the previous
part [see Eq.(\ref{SA00a})]. Now we need to calculate $S_{A_2}$, which can be obtained based
on Fig.\ \ref{DisjointTorus01} by replacing $A_1$ with $A_2$.

First, we consider the simple case that the Wilson line is in a definite representation $\hat{R}_a$.
Based on Fig.\ \ref{DisjointTorus01} (c), one can check that
\begin{equation}
\text{tr}\left(\rho_{A_2}^n\right)=\frac{Z(S^3,\hat{R}_a)^4}{Z(S^3,\hat{R}_a)^{4n}}.
\end{equation}
Therefore, one has
\begin{equation}
\frac{\text{tr}\left(\rho_{A_2}^n\right)}{\text{tr}\left(\rho_{A_2}\right)^n}
=
\frac{1}{Z(S^2\times S^1, \hat{R}_a,\hat{\overline{R}}_a)^n}\cdot\frac{Z(S^3,\hat{R}_a)^4}{Z(S^3,\hat{R}_a)^{4n}}=
\left(\mathcal{S}_{0a}\right)^{4-4n},
\end{equation}
and
\begin{equation}\label{SA200a}
S^{(n)}_{A_2}=\frac{1}{1-n} \ln\frac{\text{tr}\left(\rho_{A_2}^n\right)}{\text{tr}\left(\rho_{A_2}\right)^n}=4\ln \mathcal{S}_{0a}.\\
\end{equation}
Now we consider the general case $|\Psi\rangle=\sum_a\psi_a |\hat{R}_a\rangle$. It is straightforward to check that
\begin{align}
&
\text{tr}\left(\rho_{A_2}^n\right)=\sum_a|\psi_a|^{2n}\frac{Z(S^3,\hat{R}_a)^4}{Z(S^3,\hat{R}_a)^{4n}}
=\sum_a|\psi_a|^{2n}\mathcal{S}_{0a}^{4-4n},
\notag\\
&
\text{and}\quad \text{tr}\left(\rho_{A_2}\right)=\sum_a|\psi_a|^2=1.
\end{align}
Then one has
\begin{align}\label{SA200bb}
S^{(n)}_{A_2}&=\frac{1}{1-n} \ln\frac{\text{tr}\left(\rho_{A_2}^n\right)}{\text{tr}\left(\rho_{A_2}\right)^n}=
\frac{1}{1-n}\ln \sum_a|\psi_a|^{2n}\mathcal{S}_{0a}^{4-4n},\notag\\
\text{and}\quad S_{A_2}&=4\sum_a|\psi_a|^2\ln \mathcal{S}_{0a}-\sum_a|\psi_a|^2\ln |\psi_a|^2.
\end{align}
Then one can immediately obtain the mutual information
between $A_1$ and $A_2$ as follows
\begin{align}\label{MI200bb}
I^{(n)}_{A_1,A_2}&=
\frac{1}{1-n}\ln \sum_a|\psi_a|^{2n}d_a^{4-4n}-4\ln \mathcal{D},\notag\\
\text{and}\quad I_{A_1,A_2}&=4\sum_a|\psi_a|^2\ln d_a-4\ln \mathcal{D}-\sum_a|\psi_a|^2\ln |\psi_a|^2.
\end{align}

\subsection{Two adjacent non-contractible regions on a torus with contractible $B$}

\begin{figure}[ttt]
   \begin{center}
     \includegraphics[height=7.5cm]{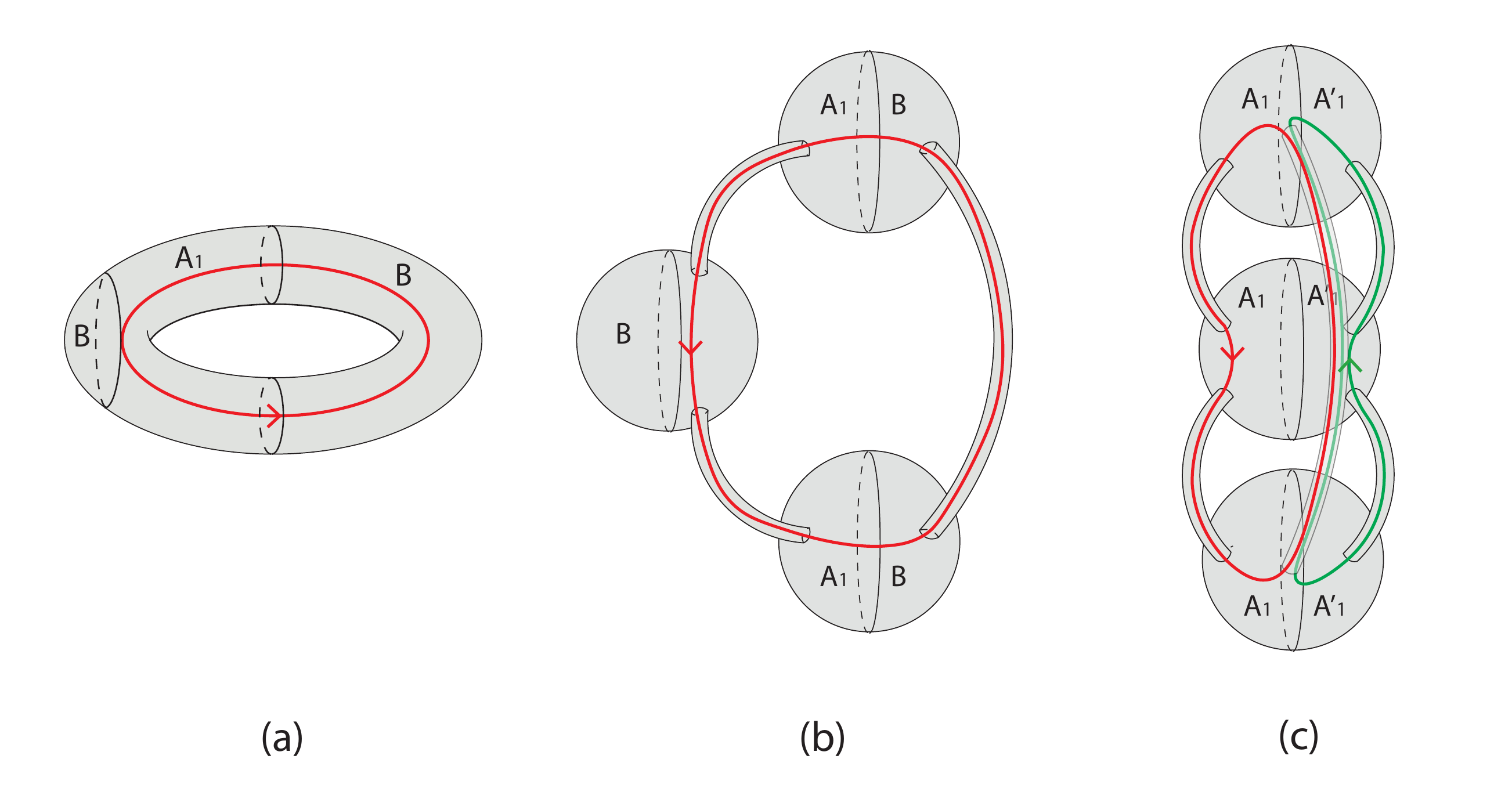}
   \end{center}
   \caption{
(a) 
Wave functional $|\Psi\rangle$, which is represented by a tripartite toroidal manifold threaded by a Wilson loop. 
(b) Deformation of $|\Psi\rangle$, without changing the topology. 
(c) $\rho_{A_1}=\text{tr}_B\left(|\Psi\rangle\langle \Psi|\right)$.
}\label{TorusTri00MIa}
\end{figure}

This case corresponds to the configuration in Fig.\ \ref{Torustri00} (a). The entanglement entropy
$S_{A_1\cup A_2}$ has already been calculated in Ref.\ \cite{Dong}, and has the following simple expression
\begin{equation}
S_{A_1\cup A_2}^{(n)}=2\ln \mathcal{S}_{00}=-2\ln \mathcal{D}.
\end{equation}
Now we need to calculate $S^{(n)}_{A_1}$ (or $S^{(n)}_{A_2}$). As shown in Fig.\ \ref{TorusTri00MIa},
for convenience, we denote the compliment part of $A_1$ as $A_1^{\complement}=B$. 
Shown in Fig.\ \ref{TorusTri00MIa} (b) is a deformation of $|\Psi\rangle$, based
on which we can obtain $\rho_{A_1}$ in Fig.\ \ref{TorusTri00MIa} (c). Then it can be checked that
\begin{equation}
\text{tr}\left(\rho_{A_1}^n\right)=\frac{Z(S^3,\hat{R}_a)^2\cdot Z(S^3)^{1-n}\cdot Z(S^3,\hat{R}_a)^n}{Z(S^3,\hat{R}_a)^{3n}}
=Z(S^3,\hat{R}_a)^{2-2n}\cdot Z(S^3)^{1-n},
\end{equation}
where we have used Eq.\ (\ref{nWilsonLine}).
Therefore, one has
\begin{equation}
\frac{\text{tr}\left(\rho_{A_1}^n\right)}
{\left(\text{tr}\rho_{A_1}\right)^n}=
\frac{1}{Z(S^2\times S^1,\hat{R}_a,\hat{\overline{R}}_a)^n}\cdot
Z(S^3,\hat{R}_a)^{2-2n}\cdot Z(S^3)^{1-n}
=\left(\mathcal{S}_{0a}\right)^{2-2n}\cdot \left(\mathcal{S}_{00}\right)^{1-n}.
\end{equation}
For the general case $|\Psi\rangle=\sum_a\psi_a |\hat{R}_a\rangle$, it is straightforward to check that
\begin{equation}
\frac{\text{tr}\left(\rho_{A_1}^n\right)}
{\left(\text{tr}\rho_{A_1}\right)^n}
=\frac{\sum_a |\psi_a|^{2n} Z(S^3,\hat{R}_a)^{2-2n}\cdot Z(S^3)^{1-n}}{\left(\sum_a|\psi_a|^2\right)^n}
=\sum_a|\psi_a|^{2n}\left(\mathcal{S}_{0a}\right)^{2-2n}\cdot \left(\mathcal{S}_{00}\right)^{1-n}.
\end{equation}
Then one has
\begin{align}
S_{A_1}^{(n)}&=\frac{1}{1-n}\ln \sum_a|\psi_a|^{2n}\left(\mathcal{S}_{0a}\right)^{2-2n} +\ln \mathcal{S}_{00},\notag\\
\text{and}\quad S_{A_1}&=2\sum_a|\psi_a|^2\ln \mathcal{S}_{0a}-\sum_a|\psi_a|^2\ln |\psi_a|^2+\ln \mathcal{S}_{00}.
\end{align}
From Fig.\ \ref{Torustri00} (a), it is straightforward to observe that $S_{A_1}^{(n)}=S_{A_2}^{(n)}$.
Then the mutual information between $A_1$ and $A_2$ has the expression
\begin{align}
I_{A_1,A_2}^{(n)}&=\frac{2}{1-n}\ln \sum_a|\psi_a|^{2n}\left(\mathcal{S}_{0a}\right)^{2-2n}=\frac{2}{1-n}\ln \sum_a|\psi_a|^{2n}d_a^{2-2n},\notag\\
\text{and}\quad I_{A_1,A_2}&=2\sum_a|\psi_a|^2\ln \mathcal{S}_{0a}-\sum_a|\psi_a|^2\ln |\psi_a|^2\notag\\
&=4\sum_a|\psi_a|^2\ln d_a-2\sum_a|\psi_a|^2\ln |\psi_a|^2-4\ln \mathcal{D}.
\end{align}

\subsection{Two disjoint non-contractible regions on a torus}

This case corresponds to the configuration in Fig.\ \ref{DisjointTorus01} (a).
$S_{A_1}^{(n)}$ and $S_{A_2}^{(n)}$ have the same expression as those in Eq.\ (\ref{SA00a}), and
$S_{A_1\cup A_2}^{(n)}$ is the same as that in Eq.\ (\ref{SA200bb}).
Therefore, it can be checked that
\begin{align}
I_{A_1,A_2}^{(n)}&=\frac{1}{1-n}\ln \frac{
\left(\sum_a|\psi_a|^{2n}\left(\mathcal{S}_{0a}\right)^{2-2n}\right)^2
}{\sum_a|\psi_a|^{2n}\left(\mathcal{S}_{0a}\right)^{4-4n}}
=\frac{1}{1-n}\ln \frac{
\left(\sum_a|\psi_a|^{2n}d_a^{2-2n}\right)^2
}{\sum_a|\psi_a|^{2n}d_a^{4-4n}},\notag\\
I_{A_1,A_2}&=-\sum_a|\psi_a|^2\ln |\psi_a|^2.
\end{align}
It is noted that although $I_{A_1,A_2}$ is independent of the quantum dimension, $I^{(n)}_{A_1,A_2}$ with $n>1$ depends on the
quantum dimension explicitly.

\section*{Acknowledgement}

XW thanks Yanxiang Shi for help with plotting.
PYC is supported by the Rutgers Center for Materials Theory group postdoc grant.
This work was supported in part by the National Science Foundation grant DMR-1455296 (XW and SR)
at the University of Illinois, and by Alfred P. Sloan foundation.

\end{document}